\documentclass{aa}  

\usepackage{graphicx}
\usepackage{txfonts,textcomp}
\usepackage{svg}
\usepackage{subcaption}
\usepackage{amsmath}
\usepackage{multirow}
\usepackage{enumitem}
\usepackage[hyphens]{url}
\usepackage{makecell}
\usepackage{hyperref}
\graphicspath{{Section_3/}{Section_4/}{Section_5/}}

\begin{document}

   \title{The cosmological analysis of X-ray cluster surveys}

   \subtitle{V. The potential of cluster counts in the $1<z<2$ range}

   \author{N. Cerardi\inst{1}
          \fnmsep\thanks{\email{nicolas.cerardi@cea.fr}}
          \and
          M. Pierre\inst{2}
          \and
          P. Valageas\inst{3}
          \and
          C. Garrel\inst{4}
          \and 
          F. Pacaud \inst{5}
          }

   \institute{Université Paris Cité, Université Paris-Saclay, CEA, CNRS,             AIM, F-91191, Gif-sur-Yvette, France
        \and 
             Université Paris-Saclay, Université Paris Cité, CEA, CNRS, AIM, 91191, Gif-sur-Yvette, France
        \and Institut de Physique Théorique, Université Paris-Saclay, CEA, CNRS, F-91191 Gif-sur-Yvette Cedex, France
        \and Max Planck Institute for Extraterrestrial Physics, Giessenbachstrasse 1, D-85748 Garching, Germany
        \and 
        Argelander-Institut für Astronomie, University of Bonn, Auf dem Hügel 71, D-53121 Bonn, Germany}

   \date{Received August 10, 2023; accepted November 17, 2023}

 \abstract
   {Cosmological studies have now entered Stage IV according to the Dark Energy Task Force (DETF) prescription. New missions (\textit{Euclid}, \textit{Rubin Observatory}, \textit{SRG/eROSITA}) will cover very large fractions of the sky with unprecedented depth. These are expected to provide the required ultimate accuracy in the dark energy (DE) equation of state (EoS), which is required for the elucidation of the origin of the acceleration of cosmic expansion. However, none of these projects have the power to systematically unveil the galaxy cluster population in the $1<z<2$ range. There therefore remains the need for an \textit{ATHENA}-like mission to run independent cosmological investigations and scrutinise the consistency between the results from the $0<z<1$ and $1<z<2$ epochs.}
   {We study the constraints on the DE EoS and on primordial non-gaussanities for typical X-ray cluster surveys executed by a generic \textit{ATHENA}-like Wide Field Imager. We focus on the impact of cluster number counts in the $1<z<2$ range.}
   {We consider two survey designs: 50 deg$^2$ at 80ks (survey A) and 200 deg$^{2}$ at 20ks (survey B). We analytically derive cluster number counts and predict the cosmological potential of the corresponding samples, A and B, by means of a Fisher analysis. We adopt an approach that forward models the observed properties of the cluster population in the redshift---count rate---hardness ratio parameter space.}
   { The achieved depth allows us to  unveil the halo mass function down to the group scale out to $z=2$. We predict the detection of thousands of clusters down to a few 10$^{13} h^{-1} M_{\odot}$, in particular 940 and 1400 clusters for surveys A and B, respectively, at $z>1$. Such samples will allow a detailed modelling of the  evolution of cluster physics along with a standalone cosmological analysis. Our results suggest that survey B has the optimal design as it provides greater statistics. Remarkably,  high-redshift clusters represent 15\% or less of the full samples but contribute at a much higher level to the cosmological accuracy:  by alleviating various degeneracies, these objects allow a significant reduction of the uncertainty on the cosmological parameters: $\Delta w_a$ is reduced by a factor of $\sim 2.3$ and $\Delta f_{NL}^{loc}$ by a factor of $\sim 3$.}
   {Inventorying the deep high-$z$ X-ray cluster population can play a crucial role in ensuring overall cosmological consistency. This will be the major aim of future new-generation \textit{ATHENA}-like missions.}


   \maketitle

\section{Introduction} \label{sec:intro}

Until very recently, observations over the entire redshift spectrum from the epoch of recombination to the present day seemed to unambiguously favour the standard cosmological model, namely $\Lambda$CDM. This assumes Gaussian random initial fluctuations that are thought to have  originated at a much earlier epoch, during the inflationary stage \citep[see][for reviews]{chen_primordial_2010, bartolo_non-Gaussianity_2004}. However, critical unknowns remain that question the validity of our current theoretical working framework: The `cosmological constant problem',  a catastrophic conflict between particle physics and cosmology, has long been identified and remains enigmatic \citep{burgess_cosmological_2013}.
Similarly, the nature of the putative dark matter (DM) has not been elucidated. Lastly, a possible tension in the Hubble--Lemaître constant  ($H_0$)  appeared after careful (supposedly bias-free) supernovae analyses in the local Universe delivered a $H_0$ value incompatible at more than 4$\sigma$ \citep[see][]{efstathiou_lockdown_2020} with that derived from the Planck CMB mission \citep{ade_planck_2011}, albeit of the same order of magnitude.  \\

Current projects and  missions (DES \citet{des_collaboration_dark_2016}, \textit{eROSITA} \citet{merloni_erosita_2012}, Euclid \citet{laureijs_euclid_2011}) have been designed to obtain deeper (and perhaps definitive) insight into the equation of state (EoS) of dark energy (DE), that is, the physical processes responsible for the acceleration of the expansion of the Universe over the last 4-6  billion years, within the scope of the so-called Dark Energy Task Force \citep{albrecht_report_2006}. In parallel, starting with JWST, the upcoming generation of instruments comes with enormous expectation:  while these missions are not necessarily primarily designed for cosmological studies, they are to provide major  breakthroughs in terms of sensitivity and angular and spectral resolution, thereby improving measurements linked to cosmology-related quantities (Rubin, \citet{ivezic_lsst_2019}; ELT, \cite{gilmozzi_european_2007}; SKA, \cite{maartens_cosmology_2015}).\\
In this context, the Advanced Telescope for High-ENergy Astrophysics\footnote{https://www.cosmos.esa.int/web/athena/home} (\textit{ATHENA}) is expected to start operations in the late 2030s. \textit{ATHENA} was selected as an L2 Mission of the ESA Cosmic Vision Program in 2014, and is designed to carry out in-depth X-ray studies of the hot and energetic Universe. The mission is to cover  a wide range of targets and topics, such as  X-ray binaries, black hole accretion and feedback, the circumgalactic medium (CGM), supernovae remnants (SNRs), active galactic nuclei (AGN), and clusters of galaxies. Two instruments, the Wide Field Imager \citep[WFI,][]{meidinger_wide_2017} and the X-ray Integral Field Unit \cite[X-IFU][]{barret_athena_2018} will share the same orientable mirror based on silicon pore optics (SPO) technology. The high sensitivity of \textit{ATHENA}, together with its narrow point-spread function (PSF), make it  ideally suited to performing deep extragalactic surveys with  the WFI (field of view 40'x40'), allowing it to detect high-redshift galaxy clusters. \\ 
\par Along with supernovae, weak shear, baryonic acoustic oscillations, and the CMB, galaxy clusters are important cosmological probes. As the most massive collapsed objects of our Universe, lying on the nodes of the cosmic web, they are tracers of both the growth of structures and of the expansion history of the Universe. The most commonly used statistics are cluster number counts \citep[e.g.][hereafter XXL paper XLVI]{bocquet_cluster_2019, garrel_xxl_2022}, cluster spatial clustering \citep[e.g.][]{marulli_xxl_2018}, and the baryon fraction \citep{mantz_cosmological_2022}. However, cluster cosmology is often challenged because the link between cluster observable properties and  mass is not straightforward. The many physical processes involving galaxies, AGN feedback, merger events, magnetic fields, and turbulence significantly complexify  the modelling of the intracluster medium (ICM) and question the validity of the hydrostatic equilibrium hypothesis. The scaling relations are difficult to estimate with precision, especially when accounting for intrinsic scatter and covariance between the parameters.  This imprecision has an impact on the determination of the cluster masses, which is a key information to test cosmological models. Independent weak lensing mass measurements may help to constrain cosmological models, but these are also affected by their own biases.  Nonetheless, many cluster surveys have already provided cosmological constraints: for example in X-rays (e.g. eBCS \cite{ebeling_rosat_2000}; XXL, \cite{pacaud_xxl_2018}, XXL paperXLVI), optical (e.g. DES, \cite{des_collaboration_dark_2020}), and S-Z (e.g. Planck, \cite{planck_collaboration_planck_2016}; SPT, \cite{bleem_galaxy_2015}), and based on catalogues containing $\sim10^3$ objects  at most. In the X-ray domain, \textit{eROSITA} will increase the number of catalogued objects by several orders of magnitude ($\sim10^5$ clusters predicted), but remains restricted to low redshifts (mostly $z<1$) and rather high masses. The \textit{ATHENA} mission will therefore provide a unique opportunity to improve our knowledge of the cluster population across a wide range of redshifts and masses.

\par The goal of the present paper is to evaluate the cosmological potential of the galaxy cluster population to be unveiled by the \textit{ATHENA} extragalactic surveys.
While the precise technical specificities of \textit{ATHENA} are still being elaborated, we believe it to be important to  quantitatively investigate  the contribution of an \textit{ATHENA}-like mission to cluster cosmology\footnote{This aspect had been totally overlooked in  \href{https://www.cosmos.esa.int/documents/400752/400864/Athena+Science+Requirement+Document/5f0c65ff-c009-02d2-cac2-123f3cbd94af}{the first science requirement document}}.  
 This question might be regarded as pointless and outdated by the time  \textit{ATHENA} is launched, but in practice, it is impossible to unambiguously anticipate the net outcome of the current \textit{eROSITA} and Euclid missions, which have been specifically designed to solve the EoS of DE: while detailed predictions exist, unexpected findings remain possible. Indeed, the current debate over the value of the Hubble--Lemaître constant clearly demonstrates that `precision cosmology' cannot be reduced to a simple accuracy problem.  Specifically, \textit{ATHENA} is expected to provide unprecedented insight into the $1<z<2$ galaxy-cluster Universe, of which \textit{eROSITA} and Euclid will only pick up the most massive entities. Future CMB experiments \citep[Simons Observatory and CMB-S4][]{the_simons_observatory_collaboration_simons_2019, abazajian_cmb-s4_2019}, although incrementally lowering the mass detection limit, will also target the high-mass end of the mass function, CMB-S4 being contemporaneous to \textit{ATHENA}.
We focus here on cosmological constraints from cluster number counts.  We follow a forward-modelling approach based on purely observational quantities. The procedure allows us to bypass the tedious rescaling of the individual cluster masses as a function of cosmology. In this approach, the statistical properties of the cluster population are summarised in three-dimensional parameter space (count rate; hardness ratio, redshift), which is analogous to the (magnitude; colour; redshift) space in the X-ray domain, which implicitly carries  information about cluster masses. The distribution of a cluster sample in this 3D observable space is sensitive to cosmology and hence constitutes an efficient summary statistics for cosmological inferences. The method we use ---ASpiX--- has been extensively tested, and has proven to be very efficient when applied to both real and simulated data \citep[][XXL paper XLVI]{clerc_cosmological_2012, clerc_cosmological_2012-1, valotti_cosmological_2018}.
We present Fisher predictions for two \textit{ATHENA}-like survey configurations, each totalling approximately 9 Ms. 
We set priors on the cosmological parameters and on the scaling relation coefficients that take into account the most up-to-date information related to these aspects, yet leaving significant freedom for cluster evolution beyond $z>1$. It is important to note that in the $1<z<2$ Universe, most of the X-ray cluster detections to date result from serendipitous discoveries, which prevents any serious cosmological analysis in this range, given the absence of reliable selection functions. In this respect, a new era is to be opened by \textit{ATHENA}. We limit our investigation to $z<2$, as Universe at higher redshifts than this limit is believed to be the realm of protoclusters (a specific \textit{ATHENA} science topic), for which the ICM  properties are very uncertain. Briefly, we consider constraints on the evolving DE EoS  and on the primordial non-gaussianities in order to provide a basis for comparison with ongoing projects.
\par The paper is organised as follows. In sect. \ref{sec:popmodel}, we present the formalism for our two cosmological tests, as well as the modelling adopted for the cluster population; this leads to the creation of the three-dimensional cluster X-ray observable diagrams (XODs) on which our cosmological analysis is based. Section \ref{sec:popdetect} describes our analytical modelling of the \textit{ATHENA} cluster selection function. In Sect. \ref{sec:results}, we perform a Fisher analysis to evaluate the cosmological potential of the \textit{ATHENA} surveys, with a special emphasis on the $1<z<2$ range and on survey B. In Section \ref{sec:discussion}, we discuss our results,  and quantify the role of the priors, that of the impact of spectroscopic versus photometric redshifts, and that of measurements errors; we compare our findings with the eROSITA forecasts. 
Summary and conclusions are presented in Sect. \ref{sec:ccl}. In Appendix \ref{app:HMFnongauss}, we present the derivation of a non-Gaussian correction of the halo mass function (HMF), and in Appendix \ref{app:highzA} we present for completeness additional results from survey A.
Throughout this paper, unless stated otherwise,  count rates are given for the \textit{ATHENA/WFI} instrument. The Fisher forecast plots display the 1$\sigma$ confidence regions in a 2D space, hence showing the 38\% limit. The tables report the 1$\sigma$ deviations (68\% limit) for a 1D distribution. Here, $\log$ is the base-10 logarithm.

\section{Cluster population modelling} \label{sec:popmodel}
\subsection{Cosmology}

\begin{table}[]
    \centering
    \caption{Fiducial cosmological parameters}
    \begin{tabular}{ccc}
        \hline
        Cosmological parameter & value & prior    \\ \hline \hline
        $\Omega_m$       & $0.315$ & -- \\
        $\sigma_8$       & $0.811$ & --  \\
        $h$              & $0.674$ & $\mathcal{N}(0.674,0.005^2)$  \\
        $\Omega_b$       & $0.0493$& $\mathcal{N}(0.0493,0.00076^2)$  \\
        $\tau$           & $0.054$ & $\delta_D(0.054)$  \\
        $n_s$            & $0.965$ & $\mathcal{N}(0.965,0.004^2)$   \\ \hline
        $w_0$            & $-1$    & -- \\
        $w_a$            & $0$     & -- \\ \hline
        $f_{NL}^{loc}$ & $0$     & -- \\ \hline
    \end{tabular}
    \tablefoot{Central values with assumed Gaussian priors are taken from \cite{planck_collaboration_planck_2020}. $\delta_D$ is the Dirac distribution and $\mathcal{N}(\mu, \sigma^2)$ is the normal distribution with mean $\mu$ and variance $\sigma^2$. An em-dash signifies that no prior is applied. Depending on the analysis, from five to seven cosmological parameters are included, only three of them with priors, namely $h$, $\Omega_b,$ and $n_s$.}
    \label{tab:cosmoparam}
\end{table}
As fiducial cosmology, we take a flat $\Lambda$CDM with the values of \cite{planck_collaboration_planck_2020}, as summarised in Table \ref{tab:cosmoparam}. This choice is driven by our focus on late-2030s cosmology: given the available constraints by that time, we want our study to benefit from external priors. In particular, the parameters $h$, $\Omega_b$, and $n_s$ are assigned Planck priors, as reported in Table \ref{tab:cosmoparam}. We discuss the impact of these priors  in section \ref{sec:discussion}. 
Cluster abundance is modelled following the halo mass function (HMF) from \cite{tinker_toward_2008}. In addition to the forecasts in the standard model, we consider two extensions of the standard cosmological model, which we develop in the following paragraphs.

\subsubsection{Dark energy} \label{sec:wzcdm}
While standard $\Lambda$CDM is sufficient to explain the current observations, the recently launched and upcoming observatories (\textit{eROSITA}, Euclid, Simons Observatory, SKA, CMB-S4 and finally \textit{ATHENA}) will provide a wealth of information with which to probe the nature of DE.
Assuming a DE EoS of the form $w=p/(\rho c^2)$, the underlying $\Lambda$CDM model fixes $w=-1$, but is subsequently generalised as in a free evolving parameter.
We model $w(a)$ following \citet{chevallier_accelerating_2001} and \cite{linder_exploring_2003}, a parametrisation referred to as the Chevallier-Polarski-Linder (CPL) :
\begin{equation}
    w(a) = w_0 + w_a(1-a)
.\end{equation}
In strict logic, the $w_0$ -- $w_a$ plane contains an exclusion zone where $w<-1$ \citep{vikman_can_2005}, but more complex theoretical models of phantom DE allow $w$ to cross the $-1$ threshold \citep[][for instance]{creminelli_effective_2009}. In the analysis case forecasting constraints on the DE EoS, both $w_0$ and $w_a$ are let free, without priors.

\subsubsection{Local primordial non-Gaussianities} \label{sec:fnlloc}

The primordial perturbation field in the standard inflationary models is described by a nearly scale-invariant and Gaussian spectrum for initial matter fluctuations. However, alternative inflationary models are expected to generate non-Gaussianities. When the latter only depend on the local Bardeen potential, they are called local primordial non-Gaussianities. The Bardeen potential is then expressed at lowest order with a quadratic term, parametrised by $f_{NL}^{CMB,loc}$:
\begin{equation}
\Phi(x) = \phi(x) + f_{NL}^{CMB,loc}
\left [ \phi(x)^2 - \langle \phi \rangle^2 \right ]
\label{eq:bardeen}
,\end{equation} 
with $\phi$ being a Gaussian field. While the curvature of the Universe is the subject of active debate \citep{efstathiou_evidence_2020, di_valentino_planck_2019}, it might be possible to distinguish between the different inflationary models attempting to explain an almost flat Universe if we can obtain precise constraints on $f_{NL}$. In particular, the detection of a strong non-Gaussian signal might constitute evidence in favour of a positive curvature (closed space) or a multi-field inflation mechanism \citep{cespedes_density_2021}. In this paper, we only investigate the case of local non-Gaussianities. We use the CMB convention, meaning that equation \ref{eq:bardeen} is imposed at $z \longrightarrow +\infty$ (similarly to \cite{valageas_mass_2010, pillepich_halo_2010}), as opposed to the LSS convention (e.g. \cite{sartoris_potential_2010}). Denoting $D_+$ the growth factor normalised for $D_+(a)/a \longrightarrow 1$ for $a \longrightarrow 0$, we have at late times $f_{NL}^{LSS} = f_{NL}^{CMB} / D_+(0) \approx f_{NL}^{CMB}/1.3$. To simplify the notation, we refer to $f_{NL}^{CMB, loc}$ using $f_{NL}$ in the following. The time-dependant linear matter density perturbation $\delta = \delta \rho / \rho$ is linked to $\Phi$ through:
\begin{equation}
\delta(\Vec{k}) = \alpha(k,z)\Phi(\Vec{k}),
\label{eq:alphaphi}
\end{equation} 
with
\begin{equation}
\alpha(k, z) = \frac{2c^2 k^2 D_+(z) T(k)}{3 \Omega_{m0} H_0^2}
.\end{equation} 

To compute the non-Gaussian mass function, we first recall the expression of $\sigma$, the variance of the smoothed density field:
\begin{equation}
    \sigma^2 = \frac{1}{2 \pi^2} \int^\infty_0 k^2 W_R(k)^2 P(k)dk ,
\end{equation}
where we introduce the window function $W_R(k)$, defined in the Fourier space as
\begin{equation}
    W_R(k) = \frac{3\left(\sin(kR)-kR\cos(kR)\right)}{(kR)^3} .
\end{equation}

To compute the mass function in the non-Gaussian case, we then follow the prescription from \cite{loverde_effects_2008}, multiplying the reference Gaussian mass function with a non-Gaussian correction $R_{NG}$ : 

\begin{equation}
R_{NG}(M,z,f_{NL}) = 1 + \frac{\delta_c}{6\nu^2} 
\left[
S_3 (\nu^2 -  1)^2 + \frac{dS_3}{dln\sigma} ( \nu^2 -  1)
\right]
,\end{equation}
where $\nu = \delta_c / \sigma$
and $S_3 = \langle \delta_R^3 \rangle / \sigma^4$. Here, $\delta_c$ is fixed to 1.686 and is the critical density contrast triggering halo collapse.

The primordial non-Gaussianity results in a non-zero skewness $\langle \delta_R^3 \rangle$ that can be computed with the following expression:

\begin{equation}
    \begin{split}
        \langle \delta_R^3 \rangle \ = \ & 6 f_{NL} \frac{8\pi^2}{(2\pi)^6}
         \int_0^{\infty}dk_1 k_1^2 \int_0^{\infty}dk_2 k_2^2 \\
         & \int_{-1}^{1}d\mu W_R(k_1)W_R(k_2)W_R(k_{12})
         \frac{\alpha(k_{12})}{\alpha(k_{1})\alpha(k_{2})}P(k_1)P(k_2)
  \end{split} 
  \label{eq:deltaR3}
,\end{equation}
where $k_{12}^2 = k_1^2 + k_2^2 + 2\mu k_1 k_2 $. 
We note that equation \ref{eq:deltaR3} differs from equation 10 in \cite{pillepich_x-ray_2012}, where the terms in $\alpha(k)$ are missing. We detail our derivation of $\langle \delta_R^3 \rangle$ in Appendix \ref{app:HMFnongauss}. We then we compute the mass function using
\begin{equation}
    \left( \frac{dN(M,z)}{dM} \right)_{NG} = 
    \left( \frac{dN(M,z)}{dM} \right)_{Tinker} R_{NG}(M,z,f_{NL}) .
\end{equation}

\subsection{Scaling relations} \label{sec:SL}

In order to create the fiducial XOD, we use the cluster scaling relation formalism. This allows us to convert cluster masses (from the HMF) into luminosity, temperature, and size. The chosen setup follows the parametrisation by \cite{pacaud_xxl_2018}: Mass is defined as $M_{500c}$, and the total mass enclosed in a spherical overdensity where the mean density is $\Delta$ times higher than the critical density of the Universe $M_{\Delta c}=4 \pi R_{\Delta c}^3\Delta\rho_c /3$, with $\Delta=500$.

\begin{table}[]
    \centering
    \caption{Fiducial cluster scaling relation parameters}
    \begin{tabular}{ccc}
        \hline
        Scaling law parameter & value & prior \\ \hline \hline
        $\alpha_{TM}$     & $0.342$  & $\mathcal{N}(0.342,0.55^2)$     \\
        $\beta_{TM}$      & $0.78$   & $\mathcal{N}(0.78,0.12^2)$      \\
        $\gamma_{TM}$     & $2./3.$  &  --     \\
        $\sigma_{TM}$     & $0.15$   & $\delta_D(0.15)$       \\ \hline
        $\alpha_{LM}$     & $-0.134$ & $\mathcal{N}(-0.134,1.93^2)$      \\
        $\beta_{LM}$      & $1.91$   & $\mathcal{N}(1.91,0.16^2)$      \\
        $\gamma_{LM}$     & $2.0$    & --      \\
        $\sigma_{LM}$     & $0.34$   & $\delta_D(0.34)$      \\ \hline
        $x_{c0}$          & $0.15$   & $\delta_D(0.15)$      \\
        $\sigma_{r_c}$    & $0.01$   & $\delta_D(0.01)$      \\\hline
    \end{tabular}
    \tablefoot{The slopes and scatters are taken from \citep{sereno_xxl_2020}, the evolutions are self-similar and the normalisations are refitted alone in order to recover the correct C1 number counts for an XXL-like survey. In total, six scaling-law parameters are included in the analysis, of which two ---the T-M and L-M evolution--- are let completely free.}
    \label{tab:SLparam}
\end{table}

To convert the cluster mass into ICM luminosity and temperature, we use the scaling relation formalism derived from XXL with HSC (Hyper Suprime-Cam) weak-lensing masses. We keep the slopes of the $T-M$ and $L-M$ fits from \citet{sereno_xxl_2020} and \citet{umetsu_weak_2020}. For both these relations, the evolution parameters are not well constrained and remain consistent with the self-similar expectations. We therefore retain the values of the self-similar case. The normalisation is then refitted alone on the XXL C1 catalogue to ensure that (i) we recover the correct C1 number counts and (ii) it compensates for the weak-lensing mass bias. In this scaling relation formalism, the temperature is computed inside a 300kpc radius, namely $T_{300kpc}$, and is given by the following equation:

\begin{equation}\label{eq:MT}
\begin{split}
\log \left( \frac{T_{300kpc}}{1 \text{keV}} \right) = \alpha_{TM} & + \beta_{TM}\log \left( 
\frac{M_{500c}}{10^{14} h^{-1}\text{M}_\odot} \right)\\
           &+ \gamma_{TM}\log(E(z)/E(z_{ref}))
\end{split}
,\end{equation}
where $E(z)=H(z)/H_0$, $z_{ref}=0.3,$ and the fitted parameters are given in Table \ref{tab:SLparam}. $L_{500}^{XXL}$, the luminosity in the [0.5 -- 2] keV energy range inside $R_{500c}$, is given by the relation:
\begin{equation}
\begin{split}
\log \left( \frac{L_{500}^{XXL}}{10^{43} \text{erg/s}} \right) = 
\alpha_{LM}& + \beta_{LM}\log
\left( 
\frac{M_{500c}}{10^{14} h^{-1}\text{M}_\odot} \right) \\
        &+ \gamma_{LM}\log(E(z)/E(z_{ref}))
\end{split}
.\end{equation}
Here, $L_{500}^{XXL}$ and $T_{300kpc}$ are the mean values of log-normal distributions with scatters $\sigma_{TM}=0.15$ $\sigma_{LM}=0.34$. Finally, we model the core radius of the cluster as in \cite{pacaud_xxl_2018}:

\begin{equation}
r_c = x_{c0} \times R_{500c}
.\end{equation}
\subsection{X-ray observable diagrams} \label{subsec:XODs}

The ASpiX analysis deals with raw observable properties of clusters: namely, cluster `count rates' (CR), that is, physical fluxes convolved with the sensitivity and energy responses of the X-ray instrument. In order to transform fluxes into CRs, we assume throughout our study that the sensitivity of the \textit{ATHENA}/WFI is equivalent to be five times that of XMM/EPIC(pn+mos1+mos2). More precisely, \textit{ATHENA}/WFI is expected to be ten times more sensitive than XMM/EPICpn at 1 keV \citep[see for instance][]{piro_multi-messenger-athena_2022}, and we further approximate that XMM/EPIC(mos1+mos2) doubles the sensitivity of XMM/EPIC.
 A cluster characterised by $z$, $T_{300kpc}$, and $L_{500}^{XXL}$ is observed at a given CR and a hardness ratio (HR). The CR is measured in the [0.5--2] keV range, and the HR is the ratio of the CRs in [1--2] keV  and in [0.5--1] keV. We use the APEC model and the AtomDB database \citep{smith_collisional_2001} to emulate the emission spectrum of the ICM, assuming a metallicity of $Z=0.3Z_\odot$. 
We use the response files of XMM/EPIC instruments to obtain the CRs in the bands of interest. 
\par %
We model cluster counts in the [0.05 -- 2] redshift range, with ten linearly spaced  bins. The CR are in the [0.0001 -- 20] c/s range, with 16 bins logarithmically spaced. The HR are in the [0.01 -- 1.2] range, with 16  logarithmically spaced bins. Hence, the z--CR--HR diagrams have a dimension  of (10, 16, 16).
\section{Detected cluster populations} \label{sec:popdetect}

\subsection{An exercise of survey design}

In this section, we describe  our assumptions to work out the cosmological potential of X-ray surveys to be carried out in the late 2030s by the WFI on board \textit{ATHENA}. These options are inspired by the  \textit{ATHENA/WFI} science requirement document; they are general enough to yield the proper order of magnitude of the impact of an  \textit{ATHENA}-like Large ESA X-ray mission on cluster cosmology.
As mentioned in the previous section, with a collecting area of the order of 1.4m$^2$ at 1keV, the \textit{ATHENA} sensitivity will be equivalent to five times that of  XMM/EPIC. Ideally, the silicon pore optics should provide an image quality of 5'' for the on-axis half energy width (HEW), but for cluster detection purposes, this is not critical.\\
\par In building the science case for \textit{ATHENA}, various deep extragalactic survey strategies were considered. 
In particular, a two-tiered survey was proposed, requiring 14×840ks+106×84 ks=20.66 Ms \citep[originally defined in][then updated]{nandra_hot_2013} mainly for the purpose of AGN science. 
The shallowest component (106 pointings at 84ks, total time $\sim 9$ M ) will cover $\sim 50$ deg$^2$, and thus seems particularly adapted to cluster discovery science at high redshift. This corresponds to a survey area similar to that of XXL \citep{pierre_xxl_2016}, but with an exposure time of 40 times longer. 
Let us call this option `survey A' and consider a second option, `survey B', that benefits from the same total observing time, but spread over 200 deg$^2$  with 20ks pointings. 
Table \ref{tab:surveys} summarises the survey parameters.\\

\begin{table*}[]
    \centering
    \caption{Survey designs modelled in this work.}
    \begin{tabular}{ccc}
        \hline
        Survey         & A  & B    \\ \hline \hline
        Area (deg$^2$) & 50 & 200  \\ \hline
        Depth (ks)   & 80 & 20   \\ \hline
        Optimal detection aperture $R_{opt}$ (arcsec) & 8. & 11. \\ \hline
        Limiting \textit{ATHENA}/WFI count rate $CR_{lim}(r<R_{opt})$ (c/s) & $7.01 \times 10^{-4}$ & $2.19 \times 10^{-3}$   \\ \hline
        Number of clusters in $0<z<2$ & 5500 & 11200   \\ \hline
        Number of clusters in $1<z<2$ & 940 & 1400 \\ \hline
    \end{tabular}
    \tablefoot{A and B share the same total exposure time ($\approx 9Ms$ for \textit{ATHENA}/WFI) . Survey A is similar to the planed generic  \textit{ATHENA}/WFI AGN survey}
    \label{tab:surveys}
\end{table*}

\subsection{Computing the selection functions}

This section presents the selection functions for both surveys A and B. Also, based on the observed signal, the selection functions can be translated  to estimate the completeness in the mass--redshift plane  given our set of scaling relations. We  assume that the telescope point spread function (PSF) is narrow enough to ensure that (i) the apparently smallest clusters remain distinguishable from point sources and (ii) it does not significantly affect the signal in the detection cell. This assumption is re-examined in section \ref{sec:discussion}. Compared to previous studies, where the detection probability was modelled using image simulations \citep{zhang_high-redshift_2020}, we set the detection limit to a given signal-to-noise ratio (S/N) in a circular aperture. This allows us to analytically derive detection probabilities but prevent the definition of samples with given levels of purity or completeness. In this prospective study, we simply assume that it is, in any case, a posteriori possible to distinguish clusters from AGN (using the optical/IR data for instance), which means that the science sample is 100\% pure.  
We take that a cluster is detected if the  S/N inside a fixed aperture radius is greater than five. This radius is fixed and optimised for the detection of clusters at $z\sim2$, as we want to test the relevance of the $1<z<2$ range for cosmology.\
\par We model the background as the sum of an unresolved X-ray component (diffuse cosmic background) plus the particle background. The diffuse background is taken from \cite{valotti_cosmological_2018}. As we consider an instrument that is five times as sensitive as XMM/EPIC(pn+mos1+mos2), we multiply the value given in \cite{valotti_cosmological_2018} by five, yielding 4.08$\times$10$^{-6}$ c/s/arcsec$^2$. For the \textit{ATHENA} particle background, we follow the technical requirement \citep[see][and references therein]{kienlin_evaluation_2018} of a flat spectrum 6.0 $\times$ 10$^{-4}$ c/keV/s/arcmin$^2$. We derive the particle background in the [0.5 -- 2] keV band to be 2.5$\times$10$^{-7}$ c/s/arcsec$^2$. The total background therefore amounts  to 4.33$\times$10$^{-6}$ c/s/arcsec$^2$; within our working assumptions, it is dominated by the diffuse background, because of the high sensitivity of our detector, and consequently the level of the particle background is expected to have little influence on the results of this study. \\
The detection radius is then determined via an iterative procedure. Below, $CR(r<R)$ is the integrated count rate up to the radius $R$, with $CR_{\infty} = CR_{r<\infty}$; and $S/N(r<R)$ is the S/N of the source inside the radius $R$.
\begin{enumerate}
    \item For a cluster of a given $M_{500c}$ and $z$, we use the forward modelling presented in section \ref{sec:popmodel} to compute its $CR_{\infty}$. 
    \item The $CR_{\infty}$ allows us to compute $CR(r<R)$, which is combined with the background to obtain $S/N(r<R)$. The optimal radius for this particular cluster is then $R_{opt} \coloneqq argmax(S/N(r<R))$.
    \item We compute $CR_{lim}(r<R_{opt})$, the solution of the equation $S/N(r<R_{opt})=5$. If $CR(r<R_{opt})$ yields a $S/N(r<R_{opt}) \leq 5.5$, we stop here and retain $R_{opt}$ and $CR_{lim}(r<R_{opt})$ as the final selection function parameters. Else, the current value of $R_{opt}$ ensures $S/N(r<R_{opt}) \textgreater 5.5$, and so we go to step 4 to iterate again to reduce $R_{opt}$ and hence detect lower-mass objects at $z$. 
    
    \item We find that the $M_{500c}$  ensures $CR(r<R_{opt}) = CR_{lim}(r<R_{opt})$ at $z$ through the assumed cosmology, scaling relations, and photon table. We then repeat steps 2 to 4 for a cluster of this new $M_{500c}$, and still at redshift $z$.
\end{enumerate}
\par
\begin{figure}
   \centering
   \includegraphics[width=8cm]{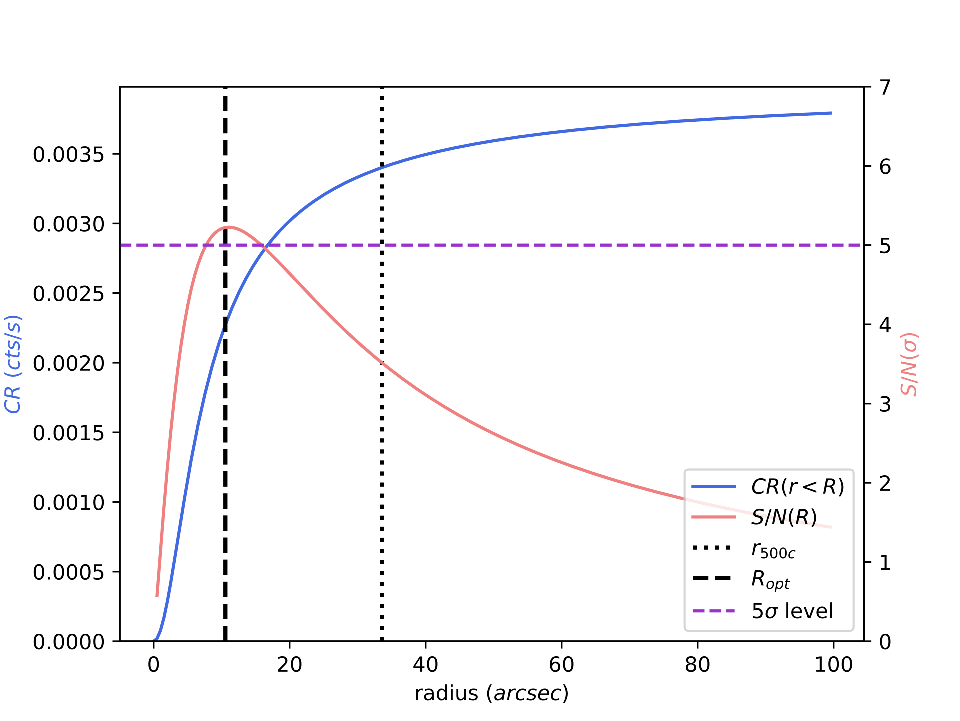}
      \caption{Integrated CR (blue) and S/N profiles (pink) of a cluster at the detection limit at $z=2$ for survey B. This cluster has a mass of $M_{500c} = 3.9 \times 10^{13} h^{-1}$ M$_\odot$, and a core radius of $r_c = 5.04"$. Its total count rate is $CR_\infty = 4.00 \times 10^{-3} c/s$. Only a reduced aperture ensures a $5\sigma$ detection. This cluster is the detection limit, and so we infer $R_{opt}=11"$ for survey B. Its count rate within the aperture is $CR(r<R_{opt}) = 2.27 \times 10^{-3} c/s$, which is just slightly above the $CR_{lim}$ quoted in Table \ref{tab:surveys}. In survey B, about 80 photons, on average, would be collected in total for this cluster.}
    \label{fig:lim_cluster_surveyB}
\end{figure}

 In our case, as we are interested in detecting high-redshift clusters, we optimise the radius for $z=2$. We run the procedure for both survey configurations independently, which results in the selection function parameters $R_{opt}$ and $CR_{lim}(r<R_{opt})$. For the purpose of illustration, figure \ref{fig:lim_cluster_surveyB} shows an integrated CR profile for a cluster at the detection limit in survey B. Table \ref{tab:surveys} summarises the detection parameters for each survey.  

\par
\begin{figure}
   \centering
   \includegraphics[width=8cm]{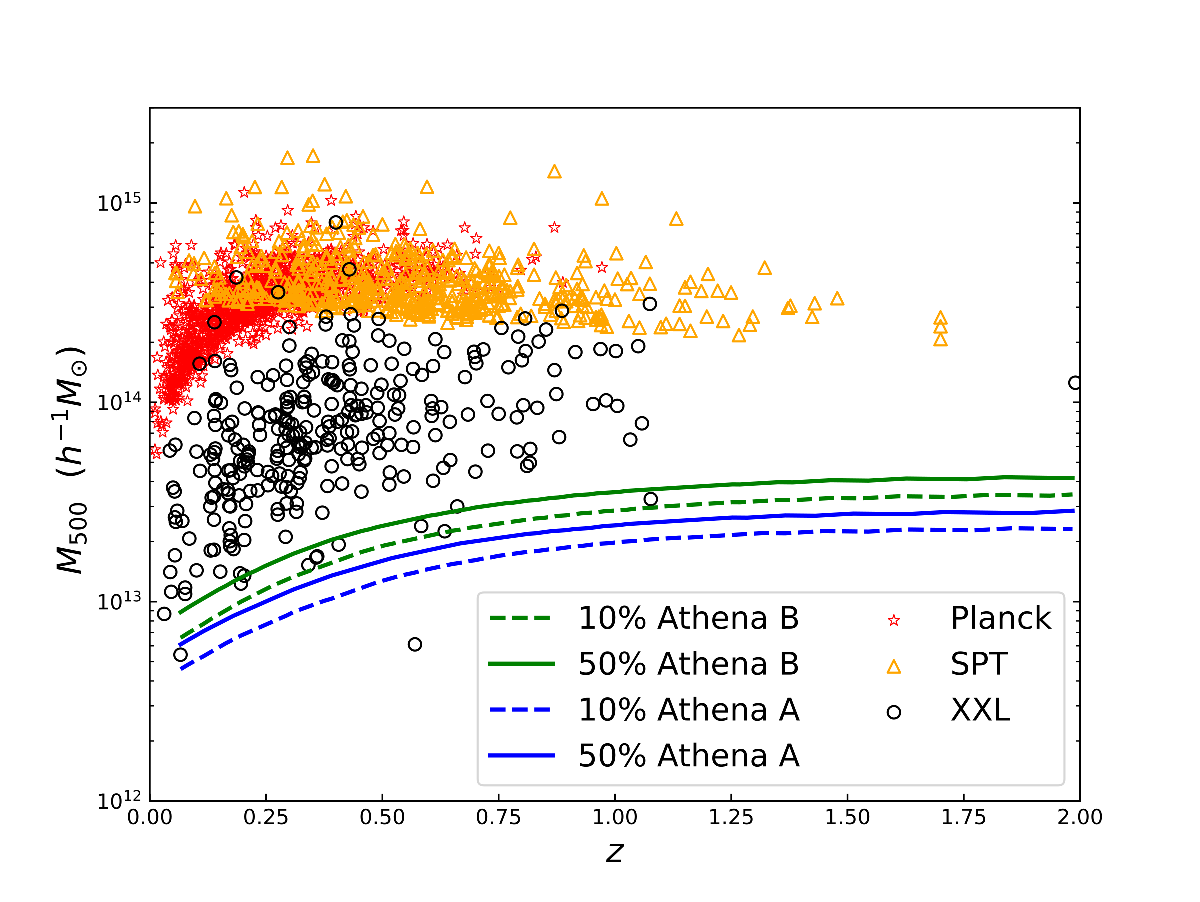}
      \caption{Mass detection limit for surveys A and B. The plain (dashed) lines represent the 50\% (10\%) detection limit in the $M_{500c}-z$ plane. A given cluster with mass above (below) a line has a detection probability of higher (lower) than  50\% in the corresponding survey design. The circles, stars, and triangles denote cluster samples from XXL \citep{adami_xxl_2018}, Planck SZ \citep{planck_collaboration_planck_2016}, and SPT \citep{bleem_galaxy_2015}, respectively.}
    \label{fig:mlimz}
\end{figure}

 The detection parameters along with the chosen scaling relations allow us to translate the  survey  selection functions in the $M_{500c} - z$ plane. Formally, an element $\Delta N (z, M_{500c}, \Omega)$ is mapped in the z--CR--HR space, taking into account the scatter in the scaling relations. Consequently, only a fraction of $\Delta N$ may exceed the survey $CR_{lim}$ and therefore be detected. This fraction is the completeness at redshift $z$ and mass $M_{500c}$, $P(\text{detection}\mid z, M_{500c})$. Combined with the slope of the HMF, this acts as an Eddington bias: there are more low-mass objects entering the sample than high-mass ones being lost. Figure \ref{fig:mlimz} represents the selection functions of surveys A and B for a detection probability of 50\%. Survey A has a lower 50\% detection limit given its deeper exposure. Interestingly, both selection functions barely evolve in the $1<z<2$ range, enabling the detection down to 2$\times10^{13}$h$^{-1}$M$_\odot$ (3.5$\times10^{13}$h$^{-1}$M$_\odot$) for survey A (survey B); this empirical behaviour is analogous to that of the S-Z detection. 
Comparing these detection limits to the expected sensitivity of CMB-S4 \citep[figure 3 in][]{raghunathan_constraining_2022}, we note that both surveys A and B will unveil lower-mass objects than those discovered by S-Z: S4-Wide (S4-Ultra Deep) will detect clusters down to $\sim10^{14}$h$^{-1}$ M$_\odot$ ($\sim 7\times10^{13}$h$^{-1}$M$_\odot$) at $z=1$ and down to $\sim 7\times10^{13}$h$^{-1}$M$_\odot$ ($\sim 5\times10^{13}$h$^{-1}$M$_\odot$) at $z=2$. The same comparison can be made with \textit{Euclid}. We can compare the completeness of AMICO \citep{bellagamba_amico_2018}, one of the retained detection algorithms for Euclid \citep{euclid_collaboration_euclid_2019}, with our derived selection functions. We take care to properly translate the $M_{DH}$ used in \citet{euclid_collaboration_euclid_2019} in our definitions, using $M_{DH}/M_{200c} \approx 1.25$, $M_{500c}/M_{200c} \approx 0.7,$ and converting from [M$_\odot$] to [h$^{-1}$M$_\odot$]. \textit{Euclid} is expected to reach 50\% completeness at $\sim 8\times10^{13}$h$^{-1}$M$_\odot$ at $z=1$ and  $\sim 2.5\times10^{14}$h$^{-1}$M$_\odot$ at $z=1.9$. 
The larger survey areas of both CMB-S4 and \textit{Euclid} ensure that can detect more clusters than our \textit{ATHENA}-like surveys, but still in a narrower mass range. 
\par However, one drawback of this detection modelling is that we miss low-mass nearby groups that have low surface brightness. In practice, such groups could be detected with an $S/N>5$ for a larger aperture, and will be readily conspicuous in the \textit{ATHENA} images  in any case. A more sophisticated selection function could adapt the size of the detection radius, but this would not significantly change our conclusions, and so we refrain from attempting to incorporate such a change.

\subsection{Cluster populations}

   \begin{figure}
   \centering
   \includegraphics[width=8cm]{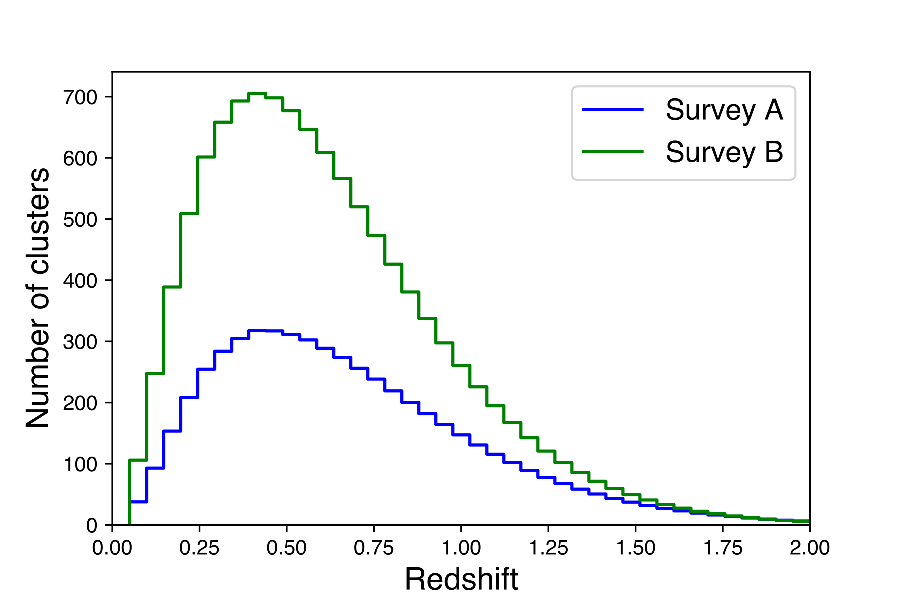}
      \caption{Number counts for surveys A and B. Survey A has a density of clusters that is about twice the density in B, but the effect of the survey area dominates and B detects twice as many clusters as A (11200 vs 5500).}
         \label{fig:dndz}
\end{figure}

\begin{figure*}
    \centering
    \begin{subfigure}{.48\textwidth}
         \centering
         \includegraphics[width=\textwidth]{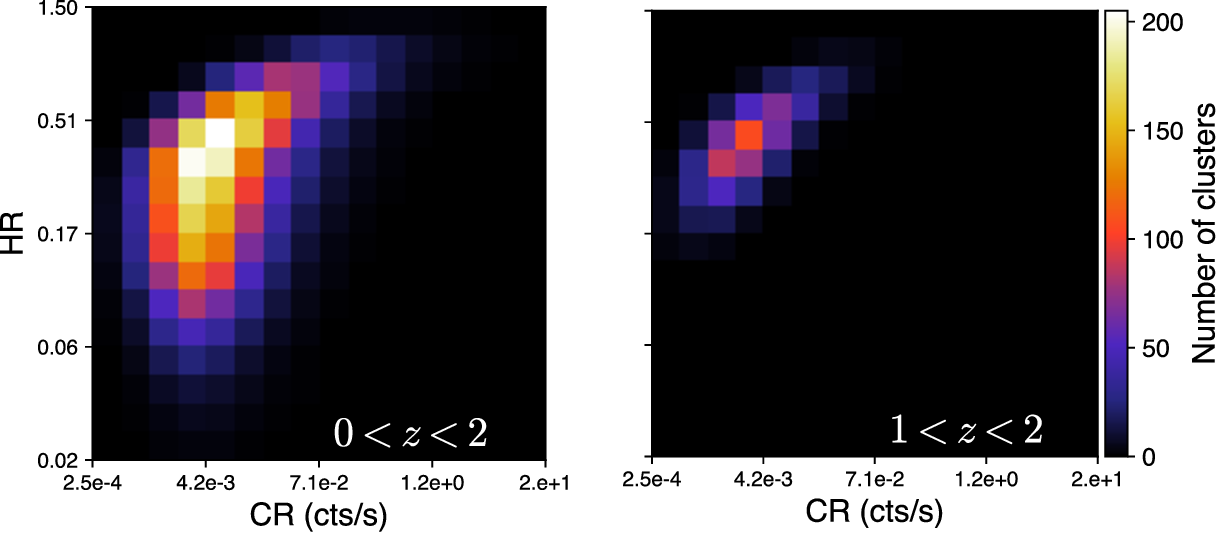}
         \caption{Survey A}
         \label{fig:xodA}
    \end{subfigure}
    \quad
    \begin{subfigure}{.48\textwidth}
         \centering
         \includegraphics[width=\textwidth]{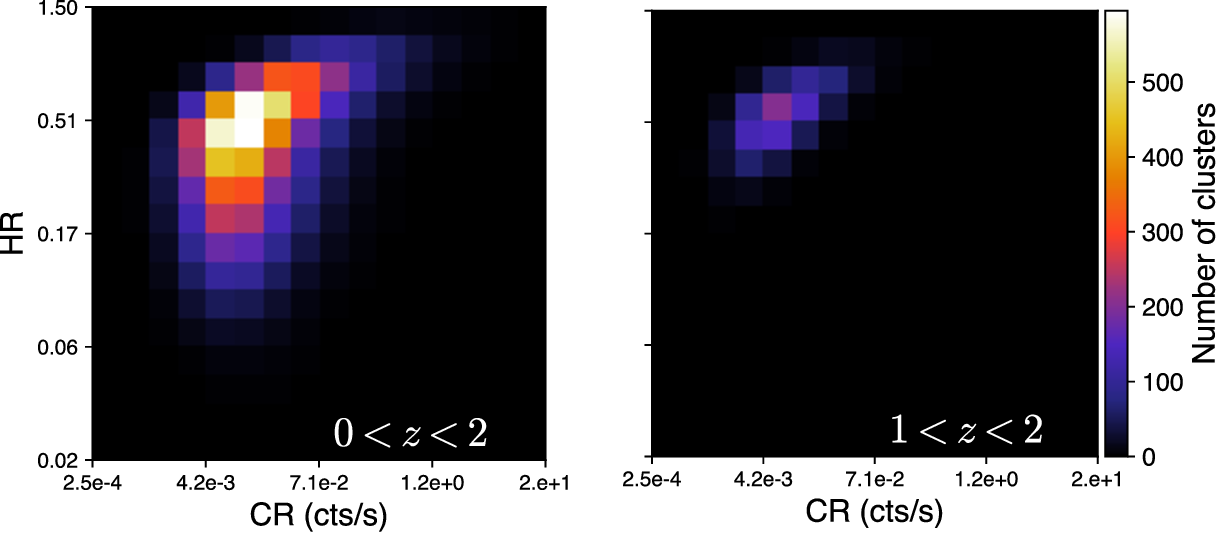}
         \caption{Survey B}
         \label{fig:xodB}
    \end{subfigure}
    \caption{Detected cluster populations in the  XOD representation. The $z$--CR--HR diagrams are integrated  over the redshift dimension  here. The colour scale  indicates the number of clusters per pixel (different range for surveys A and B).}
    \label{fig:xodAandB}
\end{figure*}
In this section, we examine the predicted detected cluster population after the selection is applied.  Figure \ref{fig:dndz} compares the redshift distribution of the two surveys.  With an exposure time four times that of survey B, survey A has a lower mass detection limit and therefore  shows a higher cluster density (about twice the density of survey B). However, survey B benefits from a four times larger area, which results in a total number of clusters of $\sim11200$,  which is twice the number found by survey A ($\sim5500$). If we focus on $z>1$, B outperforms survey A, with $\sim$50\% more clusters. The numbers are reported in Table \ref{tab:surveys}. We also point out that for both surveys, the number of clusters for $z>2$ is $\sim$20 when extrapolating both the adopted scaling relation system and the selection function.

\par The \textit{ATHENA} survey statistics are several orders of magnitude higher than current X-ray cluster samples (e.g. 365 in XXL, \citet{adami_xxl_2018}, and 1646 in X-CLASS, \citet{koulouridis_x-class_2021}), but are about an order of magnitude below the expected population of clusters in \textit{eROSITA} ($\sim$120000, see \citet{pillepich_forecasts_2018}). However, \textit{eROSITA} clusters are expected to be fewer beyond a redshift of unity (500 in eRASS:8, in only the German part; \citet{merloni_erosita_2012}). In section \ref{sec:discussion}, we compare the cosmological merits of surveys A and B with those of \textit{eROSITA}. Comparing with the XMM-Newton Distant Cluster Project \citep[XDCP,][]{fassbender_x-ray_2011}, our findings demonstrate the enormous progress expected from the Athena mission: XDCP, which covered 76.1 deg$^2$ at an average depth of 18.78 ks, was able to confirm 22 clusters at $z>0.9$.

\par The distribution of clusters in the CR--HR plane is presented in Figure \ref{fig:xodAandB}. The deeper exposure of survey A reaches lower CR and HR (so fainter and cooler), both in general ($0.05<z<2$) and in the high-redshift domain ($z>1)$. Survey A probes the HMF more deeply, but we  stress that both surveys will be able to detect clusters with $M_{500c}$ down to a few $10^{13} h^{-1} M_\odot$ out to redshift 2. This will provide invaluable information with which to follow cluster  evolution and, in particular, to investigate the scaling relations over a wide range of mass and redshift. This will be facilitated by the high number of photons collected for each cluster thanks to the long exposure time of both surveys. For example, for survey A (survey B, respectively), for a cluster at the mass detection limit, inside $R_{500c}$ we will collect 125 photons at $z\sim1$ and 100 at $z\sim2$ (85 at $z\sim1$ and 65 at $z\sim2$, respectively).  For $M_{500} = 10^{14} h^{-1} M_\odot$ at $z\sim1$, this number will rise to 3600 (900, respectively).\\

\section{Cosmological forecasts} \label{sec:results}
\subsection{Fisher formalism}
This section presents the formalism for the cosmological forecasts. We aim to use a set of $n$ observations $\Vec{x} = \{x_1, ..., x_n\}$ to constrain a set of model parameters $\Vec{\theta} = \{\theta_1, ..., \theta_p\}$.  In our case, $\Vec{x}$ is the $n$ bins of the z--CR--HR diagrams. We can rewrite the likelihood $\mathcal{L} \coloneqq P(\Vec{x} \mid \Vec{\theta})$ under the assumptions that the observations (i) are independent and (ii) follow a Poisson distribution:
\begin{equation}
    \ln \mathcal{L}(\Vec{\theta}) = \sum_i x_i \ln (\lambda_i(\theta)) - \lambda_i(\theta) - \ln (x_i!),
\end{equation}
where $\lambda_i(\theta)$ is the prediction of the model with parameters $\theta$ for the observation $i$ (in our case, bin $i$ of the diagram). The Fisher matrix is defined as:
\begin{equation}
    F_{\alpha\beta} = -\left< 
    \frac{\partial^2 \ln\mathcal{L}}
    {\partial\theta_\alpha\partial\theta_\beta}
    \right>
.\end{equation}
We can then compute each term with the following expression:
\begin{equation}
    F_{\alpha\beta} = \sum_i \frac{1}{\lambda_i}
                       \frac{\partial \lambda_i}{\partial \theta_\alpha} 
                       \frac{\partial \lambda_i}{\partial \theta_\beta}
,\end{equation}
\par which means that each bin of the z--CR--HR diagram has to be derived with respect to each parameter included in the analysis. The derivatives are numerically computed with a five-point stencil. In order to check the stability of the derivatives, they are systematically computed twice, with steps of 2\% and 5\%. We used kernel density estimators (KDEs) to project the HMF samples into the z--CR--HR in a smooth and continuous way. When using this method, the resulting
derivatives  are very stable and trustworthy. The error level on our derivatives, and therefore also on the forecasted constraints, is below 1\%. The derivatives are computed around the fiducial values listed in Table \ref{tab:cosmoparam} and Table \ref{tab:SLparam} for cosmology and for scaling relations, respectively.
\par We can then consider Gaussian priors by adding the prior matrix to the Fisher matrix. This is one of the strengths of Fisher forecasts, as one can easily test different prior configurations (see Section \ref{sec:discussion}). The major drawback is that priors must be Gaussian. The prior matrix reads as
\begin{equation}
    F_{\alpha\alpha}^{priors} = 
    \begin{cases}
    1 / \sigma_\alpha^2 \ \text{if } \theta_\alpha \text{ has a prior} \\
    0 \ \text{else}
    \end{cases}
.\end{equation}
We adopt a simple but conservative approach, and neglect the covariance terms between the parameters with priors, meaning that all non-diagonal terms are set to zero.
\subsection{Cases of forecasts}
This section presents three different cosmological forecasts. We always include in our analysis a base set of 11 parameters. Namely, we include five cosmological parameters: $\Omega_m, \sigma_8, h, \Omega_b, n_s$; and we emphasise that we take into account the normalisation, slope, and evolution parameters of the T--M and L--M scaling relations: $\alpha_{TM}, \beta_{TM}, \gamma_{TM}, \alpha_{LM}, \beta_{LM},  \gamma_{LM}$ (six parameters). Given that $\Omega_m$ and $\sigma_8$ strongly influence cluster abundances, these parameters have no prior. However, because our surveys will occur in the late 2030s, we shall benefit from robust knowledge of the priors on parameters that are harder to constrain with number counts: Planck18 priors are applied to $h, \Omega_b$, and  $n_s$. For the scaling relations, we expect the normalisation and slope to be well constrained thanks to the large cluster samples available by that time. However, we conservatively use present-day, `mild' priors from XXL on $\alpha_{TM}, \beta_{TM}, \alpha_{LM}$, and $\beta_{LM}$. We treat the evolution parameters  differently, given that surveys A and B will unveil the cluster population in a yet poorly probed mass--redshift range, likely to be very informative on $\gamma_{TM}$ and $\gamma_{LM}$. For this reason, we let the evolution parameters totally free in the analysis (i.e. no priors). In section \ref{sec:discussion}, we discuss  the impact of these choices. The cosmological forecasts are computed for the $0.05<z<2$ range, without taking into account systems at $z>2$ as (i) their number counts are very low in comparison to the full sample and (ii) beyond $z=2$, clusters are very poorly known and the HMF and scaling relations may not be reliable enough to model them.
\par This constitutes the base case of our analysis for the $\Lambda$CDM cosmology. For the $w_z$CDM scenario, we simply include $w_0$ and $w_a$ as presented in section \ref{sec:wzcdm}, without priors and keeping all the parameters and priors as defined above. For the local primordial non-Gaussianities, we include the $f_{NL}^{loc}$ as presented in section \ref{sec:fnlloc}, and set $w=-1$. Here, $w_0, w_a,$ and $f_{NL}$ are never put together in the analysis.

\begin{table*}[]
\caption{Constraints on $\Omega_m$, $\sigma_8$, $w_0$, $w_a$ and $f_{NL}$ from the Fisher analysis.}
\centering
\begin{tabular}{c|c|ccccc}
\hline
case & 
survey & $\Delta \Omega_m$ & $\Delta \sigma_8$ & $\Delta w_0$ & $\Delta w_a$ & $\Delta f_{NL}$  \\ \hline \hline
& A & 0.0065 & 0.0051 &  &  &  \\
\multirow{-2}{*}{base $\Lambda$CDM} 
& B & 0.0044 & 0.0033 &  &  & \\ \hline
& A & 0.014 & 0.0095 & 0.077 &  & \\
\multirow{-2}{*}{$w$CDM}       
& B & 0.0094 & 0.0060 & 0.054 &  & \\ \hline
& A & 0.025 & 0.015 & 0.32 & 1.25 & \\
\multirow{-2}{*}{$w_z$CDM}       
& B & 0.016 & 0.0091 & 0.22 & 0.87 & \\ \hline
& A & 0.014 & 0.038 &  &  & 470 \\
\multirow{-2}{*}{Local primordial non-Gaussianity} 
& B & 0.0094 & 0.025 &  &  & 310 \\ \hline
\end{tabular}
\tablefoot{ Nine supplementary cosmological and scaling relation parameters, not shown here, are included in the analysis and marginalised ($h$, $\Omega_b$, $n_s$, $\alpha_{TM}$, $\beta_{TM}$, $\gamma_{TM}$,
$\alpha_{LM}$, $\beta_{LM}$, $\gamma_{LM}$). The results quoted are the 1$\sigma$ uncertainties, for the full [0.05 -- 2] redshift range.}
\label{tab:constraints}
\end{table*}

\subsection{Results for the base $\Lambda$CDM}
\subsubsection{Comparison of surveys A and B on $0<z<2$}
\begin{figure}
    \centering
    \includegraphics[width=6cm]{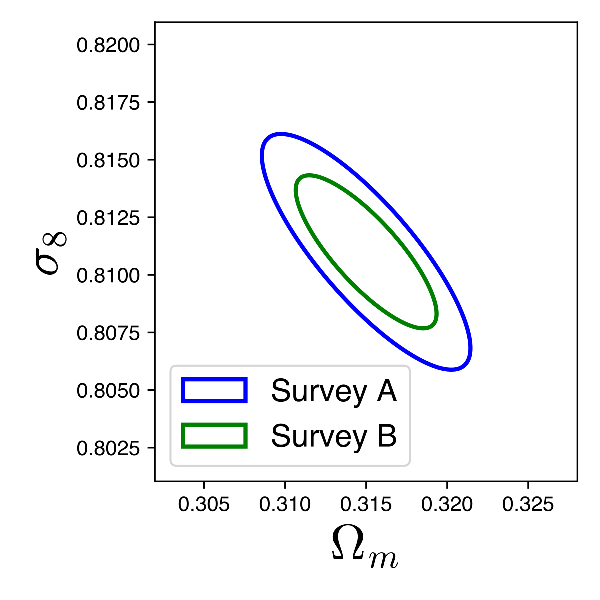}
    \caption{Comparison of the constraints on $\Omega_m, \sigma_8$ for $\Lambda$CDM provided by surveys A and B.}
    \label{fig:lcdm_AvB}
\end{figure}

As a first test case, we compare both surveys under a base  $\Lambda$CDM cosmology. Figure \ref{fig:lcdm_AvB} shows the 1$\sigma$ contours in the $\Omega_m$ -- $\sigma_8$ plane, and the corresponding uncertainties are reported in Table \ref{tab:constraints}. Survey B, with twice as many clusters, provides constraints that are approximately 1.4 times more stringent than those of survey A, which is in agreement with the fact that constraints should roughly scale according to the square root of the number of objects for detection limits that are sufficiently close. Moreover, the ellipses have the same shape and orientation, and only differ in size, which also demonstrates that the larger number of clusters in survey B is responsible for the tighter constraints. 
\subsubsection{Focus on survey B: Role of clusters at $z>1$}
\par
\begin{figure}
    \centering
    \includegraphics[width=5.5cm]{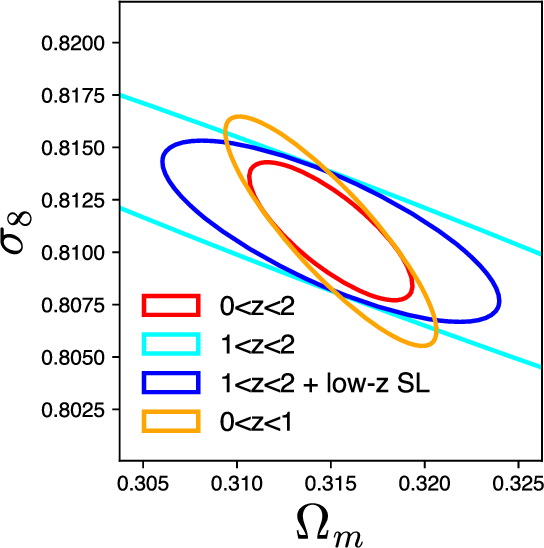}
    \caption{Survey B constraints for $\Lambda$CDM from the $0<z<1$ (orange) and $1<z<2$ (cyan) subsamples, and combination of the low-$z$ scaling relation information with the $1<z<2$ subsample (blue). The full analysis is shown in red.}
    \label{fig:lcdm_01v12_B}
\end{figure}

 We now focus on survey B, studying the cosmological information carried by clusters at $z>1$. In a first test, we separate the cluster sample in two, for $0.05<z<1$ and $1<z<2$. The corresponding constraints in the $\Omega_m$--$\sigma_8$ plane are reported in figure \ref{fig:lcdm_01v12_B}. Alone, the number counts in $1<z<2$ are limited by their modest statistics, and are less competitive than the ones in $0.05<z<1$. We then add the prior knowledge of the scaling relation parameters gained from the $0.05<z<1$ range  into the $1<z<2$ Fisher analysis. As can be seen in figure \ref{fig:lcdm_01v12_B}, the cosmological constraints are greatly improved, to a level equivalent to that of $0.05<z<1$. This indicates that survey B achieves a self-calibration of the scaling relations  at low-$z;$  this information is transferred to the cosmological modelling of the $z>1$ range. 
\par
\begin{figure}
    \centering
    \includegraphics[width=7cm]{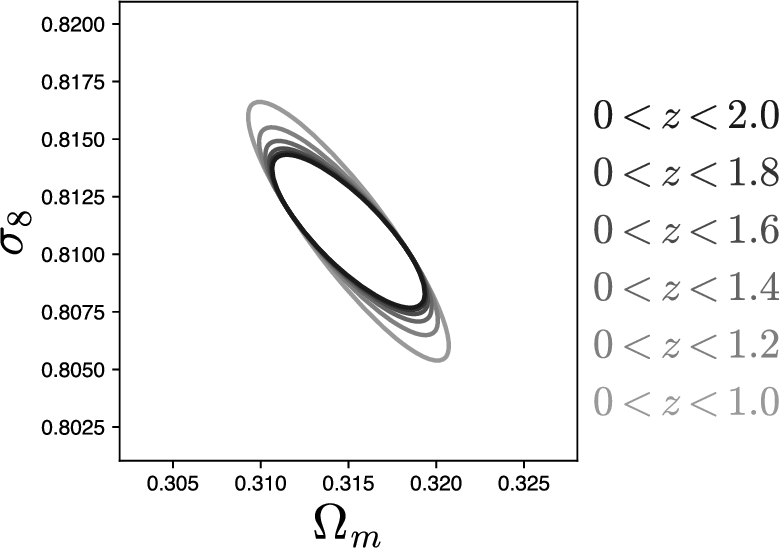}
    \caption{Impact of $z>1$ clusters in survey B on $\Omega_m, \sigma_8$ for $\Lambda$CDM. The outer and faintest ellipse corresponds to the analysis restricted to the $0.05<z<1$ range. Each successive inner and darker ellipse extends the analysis to higher redshift, up to $z=2$.}
    \label{fig:lcdm_slices_B}
\end{figure}

 We then consider the evolution of the constraints when extending the analysis range from $z=1$ to 2. Figure \ref{fig:lcdm_slices_B} shows successive confidence ellipses for redshift-truncated analyses for
survey B. The outer-most ellipse has only the lower five $z$ bins (approximately $0.05<z<1$). The inner ellipses sequentially include the five remaining bins in the Fisher analysis. We observe that high-$z$ bins induce a tilt in the ellipse, reducing the uncertainty for these parameters. Clusters at $z>1$ only represent $\sim$ 15\% of the total sample, meaning that their statistical effect is not dominant; however, they help to break the correlation between $\Omega_m$ and $\sigma_8$ within the cluster abundance analysis. $\sigma_8$ is the parameter that most benefits  from  the   high-$z$ input: constraints improve by a factor of $\sim$1.7. A similar trend is found for survey A, and the corresponding figure is reported in Appendix \ref{app:highzA}.
\par As a result, we conclude from these two tests that the self-calibration of scaling relations at low $z$ boosts the information carried by the high-$z$ subsample, allowing the degeneracies between cosmological parameters to be broken. 

\subsection{Results for $w$CDM and $w_z$CDM}
\subsubsection{Comparison of surveys A and B for the $0<z<2$ redshift range}
\begin{figure*}
   \centering
   \includegraphics[width=15cm]{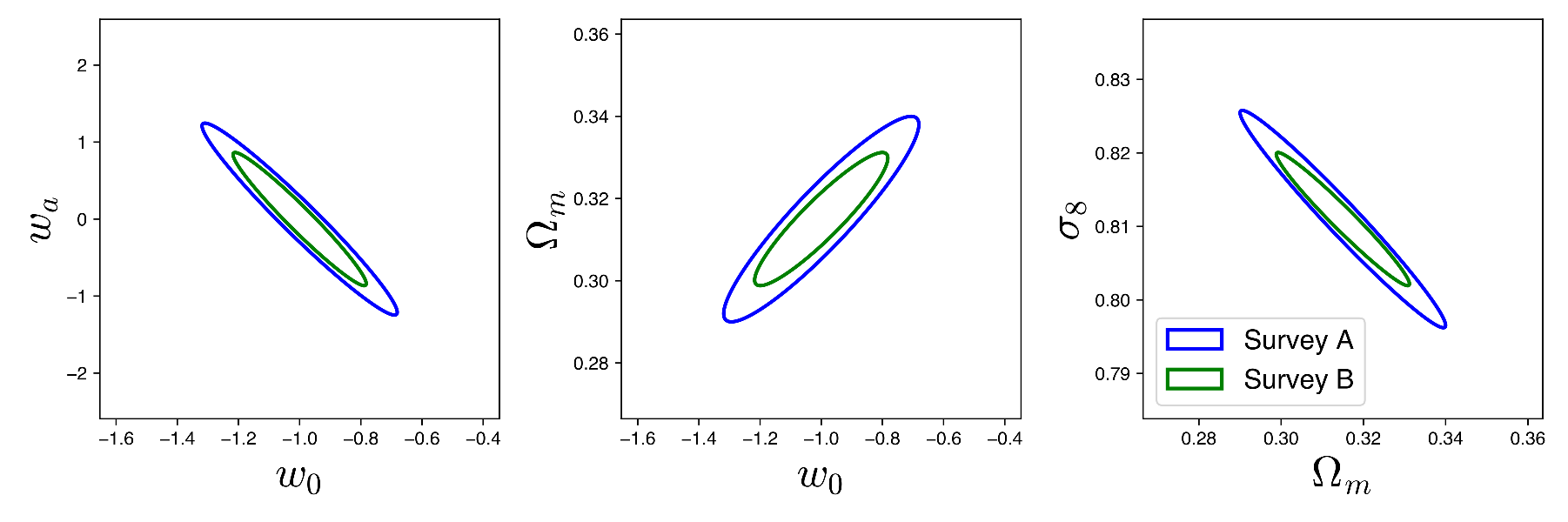}
      \caption{Comparison of the constraints on $w_0, w_a, \Omega_m$, and $\sigma_8$ for $w_z$CDM provided by  surveys A and B.}
         \label{fig:w0wa_AvB}
   \end{figure*}

We then provide forecasts for the $w$CDM and $w_z$CDM cosmology.  Constraints in the first case are reported in Table \ref{tab:constraints}. Figure \ref{fig:w0wa_AvB} focuses on the second case by showing the 1$\sigma$ contours in the $w_0$ -- $w_a$, $w_0$ -- $\Omega_m,$ and $\Omega_m$ -- $\sigma_8$ planes and the corresponding uncertainties are reported in Table \ref{tab:constraints}. Survey B is more efficient in constraining these parameters, once again thanks to its higher cluster counts. The relative improvement from survey A to survey B (tightening by a factor 1.4) is in agreement with the doubling of the cluster sample size, and we see that the ellipses share the same orientation and shape.
\subsubsection{Focus on survey B: Role of clusters at $z>1$}
\par
\begin{figure}
   \centering
   \includegraphics[width=5.5cm]{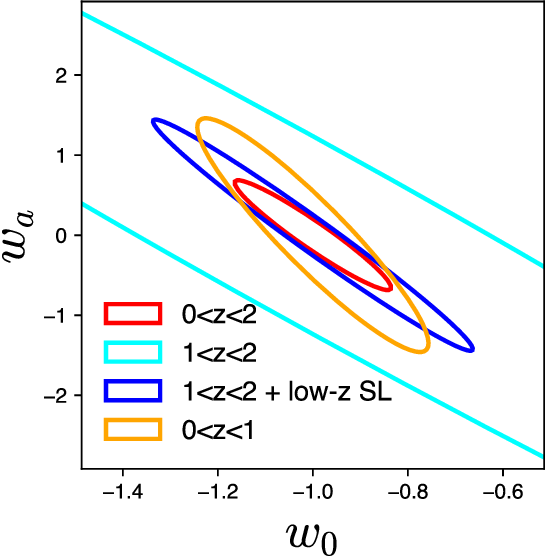}
      \caption{Survey B constraints for $w_z$CDM from the $0<z<1$ (orange) and $1<z<2$ (cyan) subsamples, and when adding the low-$z$ scaling relation information to the $1<z<2$ subsample (blue). The full analysis is shown in red.}
         \label{fig:w0wa_01v12_B}
   \end{figure}

 Figure \ref{fig:w0wa_01v12_B} compares the constraints provided by survey B  from the low-$z$ and high-$z$ subsamples in the $w_0$ -- $w_a$ plane. Again, the $1<z<2$ is far less competitive than its low$-z$ counterpart; however, the addition of the scaling relation information from $0.05<z<1$ to $1<z<2$ strongly reduces the confidence region. 
\par
\begin{figure*}
   \centering
   \includegraphics[width=15cm]{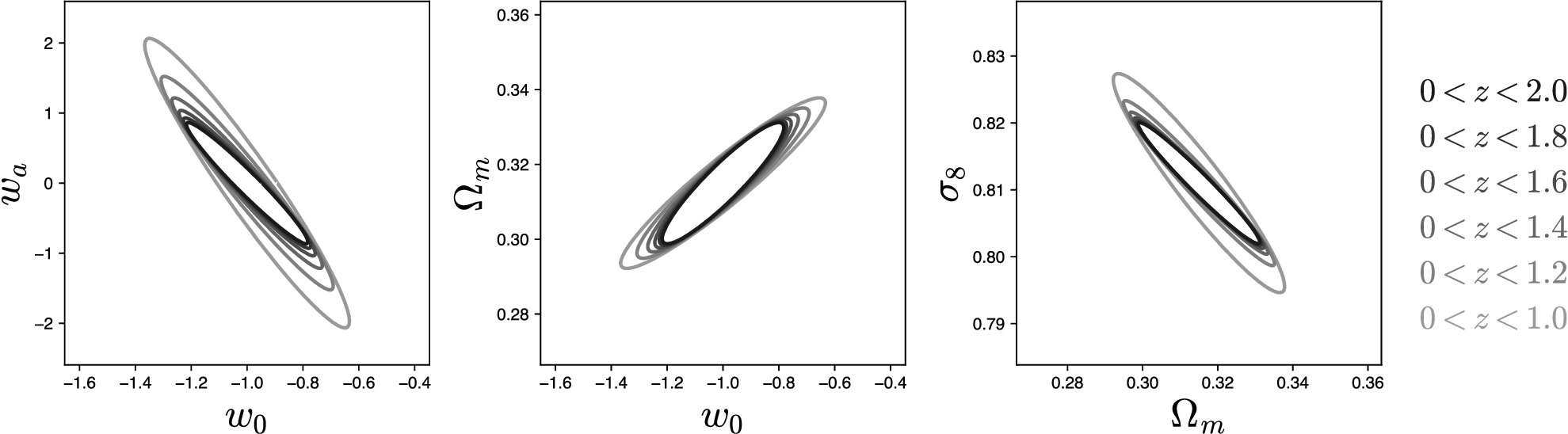}
      \caption{Impact of $z>1$ clusters in survey B on $w_0, w_a, \Omega_m$, and $\sigma_8$ for $w_z$CDM. The outer and fainter ellipse corresponds to the analysis restricted to the $0.05<z<1$ range, and the successive inner and darker ellipses extend the analysis to higher redshift, up to $z=2$.}
         \label{fig:w0wa_slices_B}
   \end{figure*}

 Figure \ref{fig:w0wa_slices_B} shows the contribution of high-$z$ clusters to the constraints resulting from survey B in the $w_z$CDM scenario. Similarly to figure \ref{fig:lcdm_slices_B}, the successive ellipses stand for the redshift-truncated analyses from approximately $0.05<z<1$ (five lower redshift bins, outermost ellipse), to $0.05<z<2$ (all ten redshift bins, innermost ellipse). We observe the same trend as for $\sigma_8$--$\Omega_m$ : clusters in high-$z$ bins induce a tilt in the ellipse, which reduces the size of the confidence region, breaking the degeneracy and improving the constraints. Importantly, because of the  degeneracy between $w_0$ and $w_a$ at $z<1$, this effect is very strong in this plane: $\Delta w_a$ shrinks by a factor of $\sim$2.3. The same trend is found in survey A, and we refer the reader to Appendix \ref{app:highzA} for the corresponding plots. 

\par While it is expected that $w_a$ is  sensitive to high-redshift systems, it is less intuitive that a small fraction of the cluster sample induces such an improvement on the cosmological constraints. This is possible thanks to the self-calibration of scaling relations at $z<1$ and the large number of clusters detected beyond $z=1$, breaking the degeneracy between $w_0$ and $w_a$.
High-redshift clusters therefore appear to be a very important component of DE investigations; their detection will require a powerful new-generation X-ray telescope such as \textit{ATHENA}.

\subsection{Results for primordial non-Gaussianities}

\begin{figure}
   \centering
   \includegraphics[width=8cm]{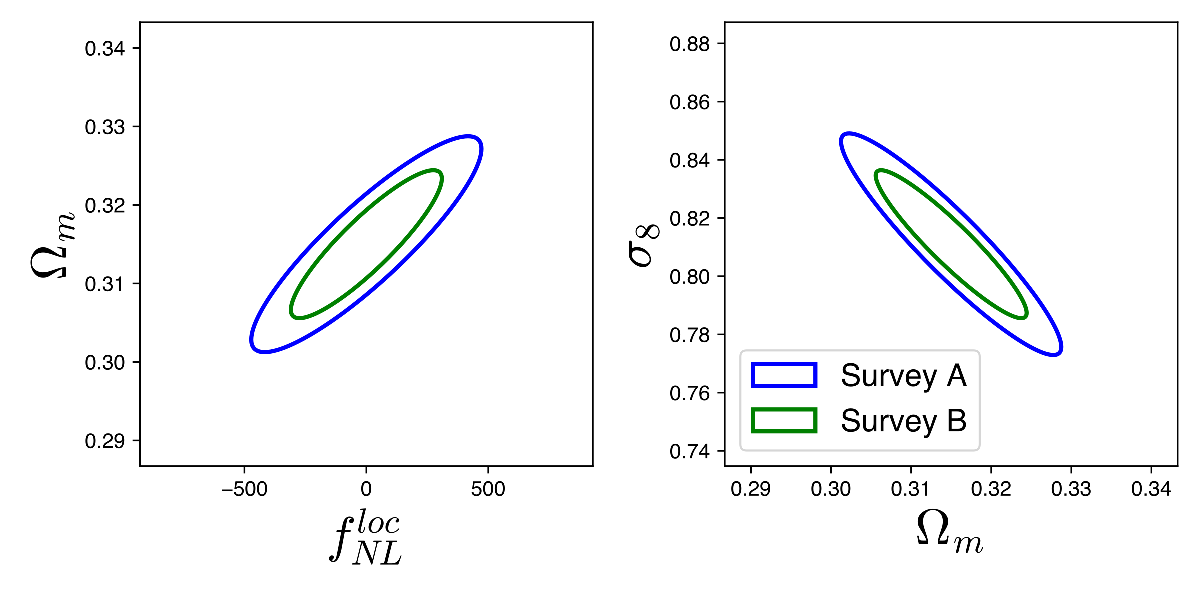}
      \caption{Comparison of constraints on $f_{NL}, \Omega_m$, and $\sigma_8$ provided by surveys A and B for local primordial non-Gaussianities (cluster counts only).}
         \label{fig:fNL_AvB}
\end{figure}

\begin{figure}
   \centering
   \includegraphics[width=8cm]{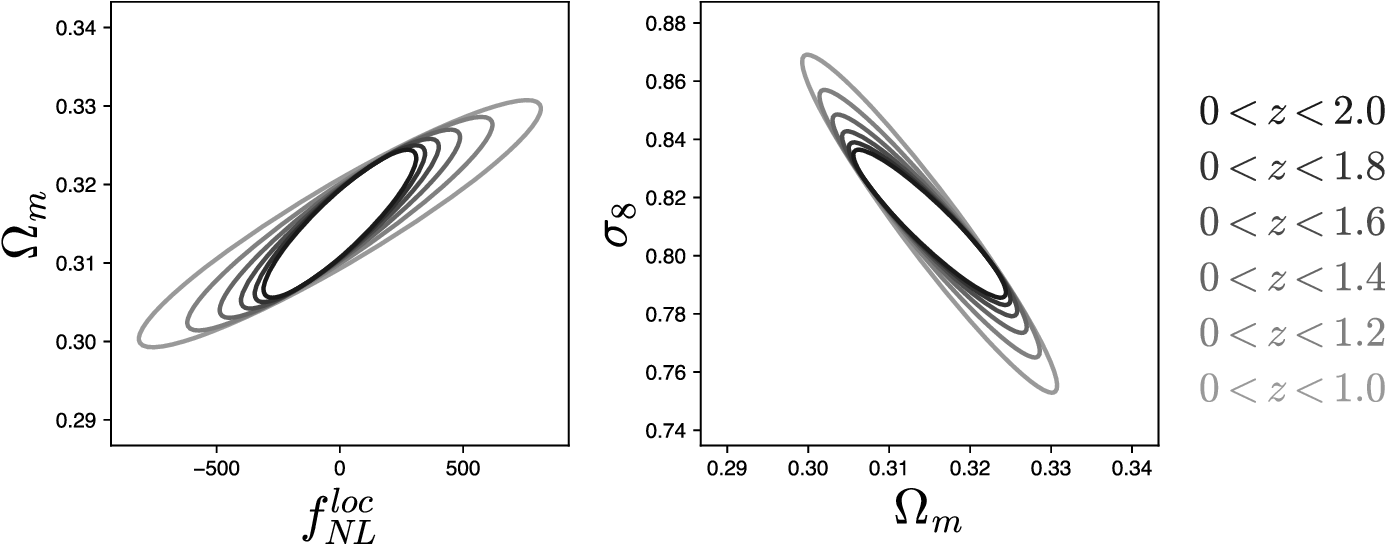}
     \caption{Impact of the $z>1$ clusters in survey B on $f_{NL}, \Omega_m$, and $\sigma_8$ for local primordial non-Gaussianities (cluster counts only).  The outer and fainter ellipse corresponds to the analysis restricted to the $0.05<z<1$ range, and the  successive inner and darker ellipses extend the analysis out to $z=2$.}
         \label{fig:fNL_slices_B}
\end{figure}

In this section, we focus on primordial non-Gaussianities. Figure \ref{fig:fNL_AvB} shows the 1$\sigma$ contours in the  $f_{NL}^{loc}$ -- $\Omega_m$ and $\Omega_m$ -- $\sigma_8$ planes; the corresponding uncertainties are reported in Table \ref{tab:constraints}. Again, survey B yields better constraints than survey A thanks to the number counts. However, we note that cluster counts alone from these surveys do not yield competitive constraints on $f_{NL}^{loc}$. We still show the contribution of high-$z$ clusters in figure \ref{fig:fNL_slices_B}. Here as well, including clusters from $z>1$ to $z\sim2$ allows us to improve the constraints on $f_{NL}^{loc}$ by a factor of $\sim 2.6$. 

\subsection{Constraints on the growth of structures}

\begin{figure}
   \centering
   \includegraphics[width=8cm]{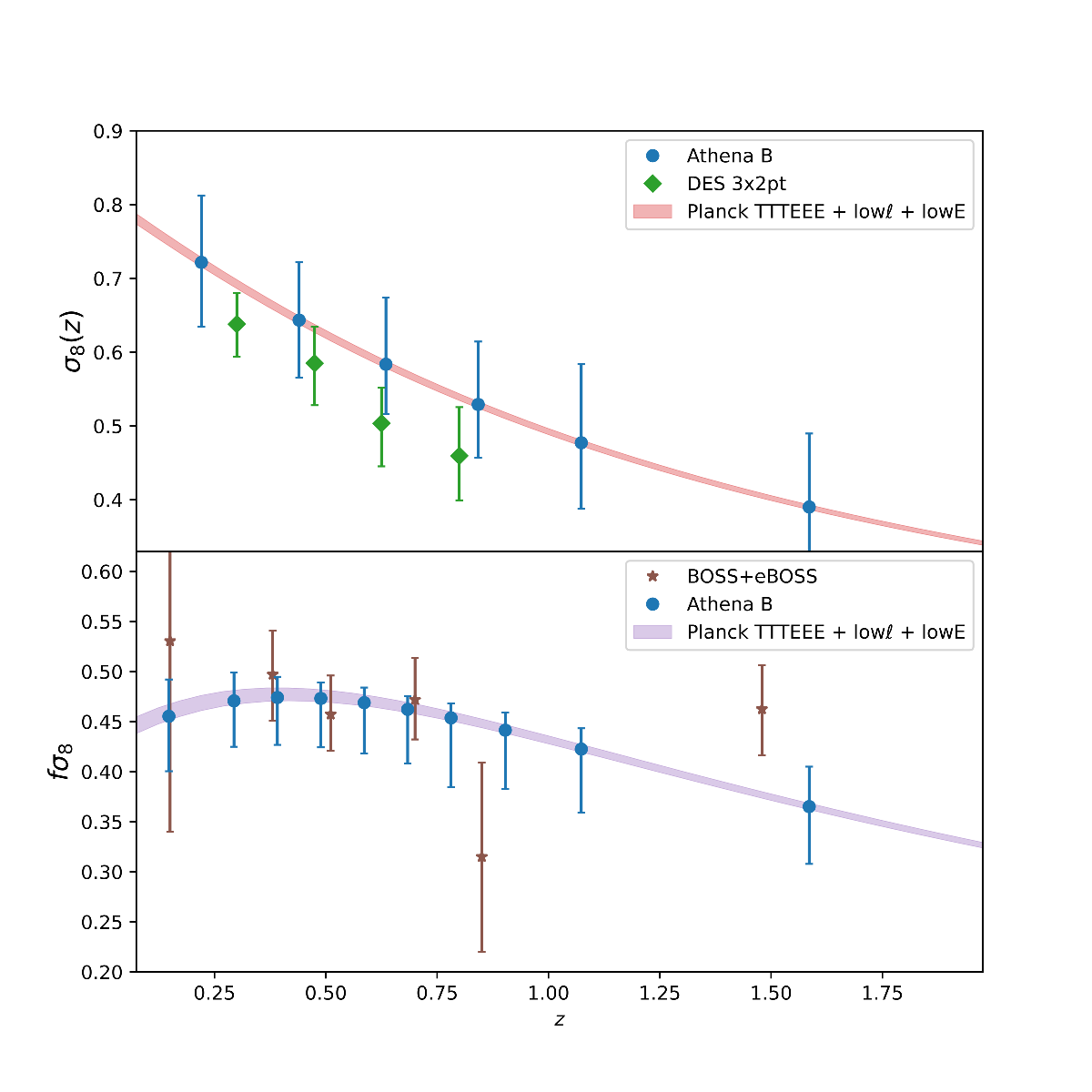}
     \caption{Constraints on $\sigma_8(z)$ (top panel) and $f\sigma_8$ (bottom panel) from independent redshift bins in survey B. For comparison, we show current constraints from DES 3x2pt \citep{des_collaboration_dark_2023} and eBOSS \citep{alam_completed_2021}.}
         \label{fig:fsigma8}
\end{figure}

Finally, we study the constraints on the growth of structures from independent cluster subsamples in  different redshift bins. We compute constraints on the time-dependent amplitude of the fluctuations, $\sigma_8 (z) = \sigma_8 D(z)$, as well as on the growth rate $f\sigma_8 = \sigma_8 (z) \times \partial\ln{D}/\partial\ln{a}$. Although cluster number counts do not directly measure these quantities, they constrain $\sigma_8$ and other primary parameters, and therefore also the growth amplitude and rate through the assumed cosmological model. In this section, we present survey B constraints on $\sigma_8(z)$ and $f\sigma_8$ assuming a $\Lambda$CDM cosmology. For $\sigma_8(z),$ we rebin the XOD along the redshift dimension in order to have $\sim$1000 clusters per bin. For $f\sigma_8$, we further regroup the bins at $z<1$ to have $\sim$2000 clusters in each. This allows us to compare the error bars obtained with current probes measuring these quantities, as shown in figure \ref{fig:fsigma8}. We compare constraints on $f\sigma_8$ with measurements from eBOSS \citep{alam_completed_2021}, and constraints on $\sigma_8(z)$ with DES 3x2pt alone \citep{des_collaboration_dark_2023}. Both quantities are also compared with the uncertainties from \citep{planck_collaboration_planck_2020}. We find that \textit{ATHENA} will deliver challenging constraints with respect to other late-time probes, and, importantly, will be able to constrain the growth of structures at $z>1$. However, this constraint using independent cluster $z$ subsamples does not exploit the full potential of \textit{ATHENA}. Indeed, we recall that this is only achieved when combining the complete redshift range $0<z<2$, as shown in the previous subsections.

\section{Discussion} \label{sec:discussion}

Our study shows that X-ray clusters, detected out to $z\sim2$ thanks to \textit{ATHENA}, can play an independent and critical role in cosmological studies. We now analyse the impact of our working hypotheses and compare our findings with predictions from other major cluster surveys.

\subsection{Uncertainties on the number counts}

Throughout this work, we neglect the effect of the PSF on cluster detection, which may not be entirely justified for high-$z$ low-mass clusters. For clusters with a small apparent size  of $z \sim2$, PSF blurring could lead to the loss of a significant fraction of the flux in the detection aperture, hence lowering  their S/N.
\par Following the prescription of the ESA online resources\footnote{https://www.cosmos.esa.int/web/athena/resources-by-esa} for \textit{ATHENA}, we model the PSF in the form of a modified pseudo-Voigt function:

\begin{equation}
    F(r) = (1-\alpha)exp\left(-\frac{r^2}{2\sigma_g^2}\right)
          + \alpha \left( 1 + \frac{r^2}{\sigma_c^2}\right)^{-\beta}
.\end{equation}

For an on-axis 1 keV source, we have HEW=5" and FWHM=3", and obtain: $\sigma_g=1.28$" and $\sigma_c=1.96$". These values can be compared to a cluster  at the detection limit with $r_c \sim 5$" (see figure \ref{fig:lim_cluster_surveyB}); point sources will be about three times smaller than this cluster. To estimate the flux loss in the detection aperture, we convolve a King profile with the PSF.
For survey B at the $z=2$ detection limit, less than 10\% of the signal is spread outside the aperture radius, meaning that the S/N will  only be decreased by about 5\%. As we retained a S/N threshold of 5, clusters at the detection limit keep a reasonable significance. We conclude that  neglecting the PSF ---which means that systems that would
be undetectable otherwise are included--- does not significantly affect our sample.
\par Moreover, a larger PSF would not invalidate the cluster counts presented in this study. The methodology adopted in section \ref{sec:popdetect} is designed to provide a more realistic selection function than a simple flux cut, but we anticipate that sophisticated two-step detection algorithms involving a convolution by the PSF and a detection probability computed in the [flux, apparent-size] parameter space \citep[e.g.][]{pacaud_xmm_2006} will be applied to the \textit{ATHENA} survey data.
For the time being, given the current uncertainties on the final  \textit{ATHENA} technical characteristics (PSF, total effective area and field of view of the WFI), we find it unnecessary to run more sophisticated calculations. 
We simply note that a larger PSF would be more problematic for AGN detection in very deep surveys, as this would raise the confusion limit.

\par For comparison, an alternative independent derivation of \textit{ATHENA} cluster counts can be found in \cite{zhang_high-redshift_2020}. Clusters follow the scaling relations from \cite{reichert_observational_2011} in the  $\Lambda CDM$ cosmology, assuming 
$\Omega_m = 0.27$, $\Omega_\Lambda = 0.73$, and $H_0 = 70 $km$/$s$/$Mpc.  Cluster detection is modelled from dedicated \textit{ATHENA} simulated images containing  both AGN and clusters (the latter being described by a beta model). Two values for the PSF HEW are considered (5'' and 10''). The cluster detection method is analogous to that of \citet{pacaud_xmm_2006}, but incorporates an additional constraint on the contribution from a possible central AGN. 
This yields $\sim 600$ clusters for the $1<z<2$ range.
The estimate presented in our paper appears to be comparable to the 10'' HEW and pessimistic detection case of \cite{zhang_high-redshift_2020}. We are  therefore  confident that our cosmological predictions constitute a conservative solution, also considering the still rather large freedom on the choice of the many parameters entering the current analysis: scaling relation coefficients, cosmology,  X-ray background properties, \textit{ATHENA} sensitivity, and PSF.
All in all, we recall that, up to a certain point, the cosmological constraints roughly scale as the square root of the number of clusters; therefore, our main conclusion as to the cosmological relevance of the $1<z<2$ range with respect to the low-z Universe should not be  affected by the particular choices made in this analysis.

\subsection{Comparison with \textit{eROSITA}}
In this section, we compare our \textit{ATHENA} predictions with cosmological forecasts for \textit{eROSITA}, regarding both  local primordial non-Gaussianities and the DE EoS (\citet{pillepich_x-ray_2012}, hereafter PP12; \citet{pillepich_forecasts_2018}, hereafter PR18), using cluster number counts only. 
The proposed comparison is an excellent way to clarify the respective roles of coverage and cosmic depth and to quantitatively explore the cluster mass ranges covered  by these two X-ray missions.
When trying to closely reproduce the PP12 and PR18 assumptions, cluster physics modelling, and selection function, we are able to derive comparable cosmological constraints. However, as our working setup (sections \ref{sec:popmodel} and \ref{sec:popdetect}) significantly differs from those of PP12 and  PP18 (different scaling relations and priors, meaning that the expected number count and constraint forecasts are not directly comparable), we applied our procedure to the \textit{eROSITA} all-sky survey definitions in order to compare both X-ray missions on the same basis.  
\par Following PP12 and PR18, we assume that \textit{eROSITA} has a sensitivity equivalent to the \textit{XMM/EPIC} instruments, and consider a 27 145 deg$^2$ survey (66\% of all-sky) with 1.6ks depth. Globally, this means that we are comparing with \textit{ATHENA}, which is five times more sensitive and has exposures some 50 and 12 times longer. Still following PP12 and PR18, we select clusters with at least 50 counts in total, meaning that $CR_{\infty, lim} = 50 / 1600 = 3.125 \times 10^{-2}$ c/s, and we refer to this survey as eRASS:8 All-Sky. Within our framework, eRASS:8 All-Sky recovers $\sim$98000 clusters, of which 600 are at $z>1$, with masses significantly higher than the ones detected through \textit{ATHENA}. The \textit{ATHENA} surveys will therefore be complementary to \textit{eROSITA} as they will systematically unveil a population of X-ray clusters undetected otherwise. \par To provide Fisher forecasts, we firstly only use the redshift and CR information, as done in PP12 and PR18, and then add an extra dimension with the HR. The same binning is applied as in the previous section, although the CR and HR windows are adapted for the eRASS population. The derived Fisher forecasts are reported in Table \ref{tab:discuss_eRASS}, where they are compared with the \textit{ATHENA} survey B. For the DE EoS, eRASS:8 All-Sky z--CR outperforms \textit{ATHENA} B by only 20\% on $w_a$, and 25\% on $w_0$: this is surprising as eRASS:8 All-Sky has about ten times more clusters. This shows the higher informative content of high-redshift clusters with respect to low-redshift when constraining the DE EoS. However, for $f_{NL}$, eRASS:8 All-Sky z--CR finds constraints that are  approximately three times more precise than those provided by \textit{ATHENA} B. We note that the eRASS:8 All-Sky z--CR constraints are also about three times more precise than in PR12. This may be caused by our scaling relation formalism when adapted to low-mass samples, which predicts more detected clusters at $z>1$ than in this latter study. For completeness, we also report constraints for eRASS:8 All-Sky z--CR--HR, but the HR in this low-exposure surveys is likely to be excessively affected by measurement errors.
We finally stress that this comparison between \textit{eROSITA} and \textit{ATHENA} is unfair towards the latter, given that eRASS:8 All-Sky is a $\sim$ 50Ms survey, while \textit{ATHENA} B is only $\sim$ 10Ms.
\begin{table}[]
\caption{Comparison of \textit{ATHENA} B cosmological potential with eRASS:8}
\centering
\begin{tabular}{c|c|ccc}
\hline
Case & 
Param. & \textit{ATHENA} B & \makecell{eRASS:8 \\ All-Sky \\ z--CR} &  \makecell{eRASS:8 \\ All-Sky \\ z--CR--HR} \\ \hline \hline
& $\Delta \Omega_m$ & 0.0094 & 0.0042 & 0.0025 \\
& $\Delta \sigma_8$ & 0.0060 & 0.0020 & 0.0013 \\
\multirow{-3}{*}{$w$CDM}   
& $\Delta w_0$      & 0.054 & 0.033 & 0.024\\ \hline

& $\Delta \Omega_m$ & 0.016 & 0.0097 & 0.0039 \\
& $\Delta \sigma_8$ & 0.0091 & 0.0038 & 0.0026 \\
& $\Delta w_0$      & 0.22 & 0.16 & 0.079 \\
\multirow{-4}{*}{$w_z$CDM}       
& $\Delta w_a$      & 0.87 & 0.69 & 0.40 \\ \hline
& $\Delta \Omega_m$ & 0.0094 & 0.0043 & 0.0032 \\
& $\Delta \sigma_8$ & 0.025 & 0.011 & 0.0092\\
\multirow{-2}{*}{\shortstack[c]{Primordial \\ non Gauss.}} 
& $\Delta f_{NL}$ & 310 & 130 & 110 \\ \hline
\end{tabular}

\label{tab:discuss_eRASS}
\end{table}
\par As an all-sky \textit{eROSITA} survey with 1.6ks exposure is now an optimistic perspective given the mission status, we can expect that (i) only the German part of the sky will be accessible and (ii) only four out of eight scans will be carried out. Taking the assumption that, in eRASS:8 All-Sky, 50 counts yield an S/N of 5, then about 40 counts are needed in eRASS:4 Half-Sky in order to reach the same S/N of 5. With our framework, we expect eRASS:4 Half-Sky to detect $\sim$30000 clusters: this will significantly impact the obtained constraints.

\par
\begin{figure}
   \centering
   \includegraphics[width=8cm]{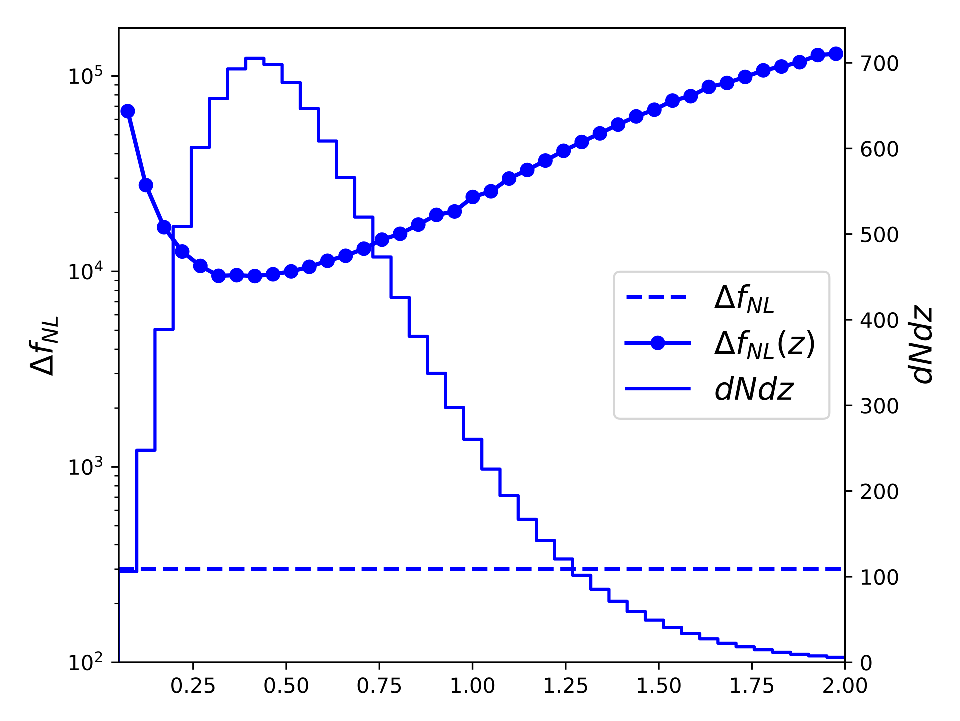}
      \caption{Constraints on $f_{NL}^{loc}$ from individual redshift bins in survey B (points). The dotted line shows the constraint of all redshift bins combined together, and the histogram represents the number counts.}
         \label{fig:fNLz_B}
\end{figure}

 Angular clustering is a key statistic for local primordial non-Gaussianities. While this paper focuses on cluster number counts, our comparison with PR12 suggests how spatial clustering would improve the constraints from surveys A and B. In the results presented in section \ref{sec:results}, we can take a closer look at the contribution from the individual redshift bins (Fig. \ref{fig:fNLz_B}) in survey B. The bins that lead to the tightest constraints correspond to the peak of the number count distribution, around  $z\sim0.4$. 
In figure 10 of PR12, the authors find that the most relevant bins are at $z\sim0.8$, which is far past their number count peak at $z\sim0.2$, but their constraints are dominated by the cluster-clustering analysis. As both our surveys detect many high-$z$ objects, cluster clustering may be a very promising probe with which to improve our forecasts.\\

\subsection{Measurement errors}

We neglected in our analysis the measurement errors on CR and HR. In practice, such errors can negatively impact the constraint as they blur the X-ray observable diagrams and therefore dilute the cosmological information. Here, we consider simple error models for both CR and HR. For the normalisation of the error models, we take the values from XXL paper XLVI and we rescale according to the observed CR of the object and the survey depth $T_{exp}$ :
\begin{equation}
    \%{err_{CR}} = 0.016 CR^{-0.5} \sqrt{\frac{10ks}{T_{exp}}}
,\end{equation}
\begin{equation}
    \%{err_{HR}} = 0.030 CR^{-0.5} \sqrt{\frac{10ks}{T_{exp}}}
.\end{equation}
\begin{figure}
   \centering
   \includegraphics[width=8cm]{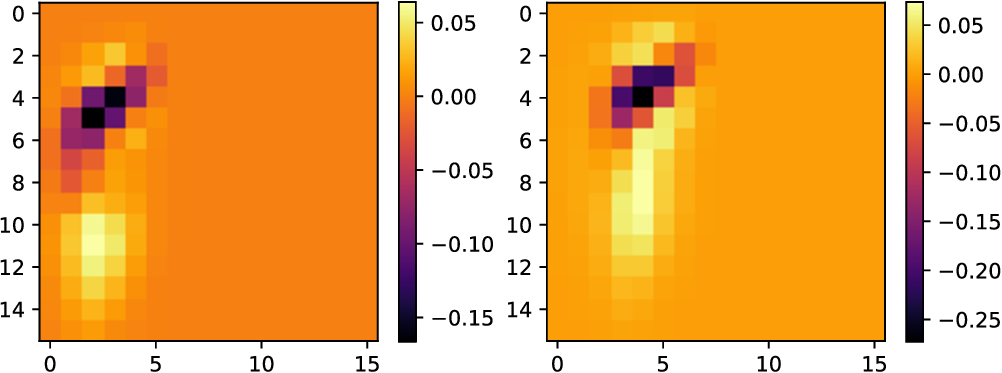}
      \caption{Difference between fiducial diagrams for surveys A (left) and B (right) with and without the measurement errors. The differences are rescaled as a fraction of the maximum value of the fiducial without errors.}
         \label{fig:errors}
\end{figure}
\par We then filter the diagrams with bivariate Gaussian convolution kernels, the scatter of which along the CR and HR dimensions depends on each bin. Figure \ref{fig:errors} shows the effect of this treatment. In practice, we observe that for surveys A and B, the error on the CR is very small ---including at low CR--- thanks to the sensitivity of WFI and the long exposure time. For the HR, there is a moderate effect on the low-CR end of the diagram. We also observe that, logically, survey A is less impacted than B by the measurement errors.

\begin{table*}[]
\centering
\caption{Impact of measurement errors on the constraints provided by surveys A and B, and eRASS:8.}
\begin{tabular}{c|c|cccccc}
\hline
Case & 
Param. & \makecell{Survey A \\ no errors} & \makecell{Survey A \\ with errors} & \makecell{Survey B \\ no errors} & \makecell{Survey B \\ with errors}  & \makecell{eRASS:8 \\ no errors}  & \makecell{eRASS:8 \\ with errors} \\ \hline \hline
& $\Delta \Omega_m$ & 0.014 & 0.015 & 0.0094 & 0.0096 & 0.0025 & 0.0026\\
& $\Delta \sigma_8$ & 0.0095 & 0.0097 & 0.0060 & 0.0061 & 0.0013 & 0.0014 \\
\multirow{-3}{*}{$w$CDM}   
& $\Delta w_0$      & 0.077 & 0.078 & 0.054 & 0.055 & 0.024 & 0.025\\ \hline

& $\Delta \Omega_m$ & 0.025 & 0.26 & 0.016 & 0.017 & 0.0039 & 0.0044\\
& $\Delta \sigma_8$ & 0.015 & 0.015 & 0.0091 & 0.0094 & 0.0016 & 0.0018\\
& $\Delta w_0$      & 0.32 & 0.33 & 0.22 & 0.23 & 0.079 & 0.091\\
\multirow{-4}{*}{$w_z$CDM}       
& $\Delta w_a$      & 1.25 & 1.27 & 0.87 & 0.89 & 0.40 & 0.45\\ \hline

& $\Delta \Omega_m$ & 0.014 & 0.014 & 0.0094 & 0.0095 & 0.0032 & 0.0033\\
& $\Delta \sigma_8$ & 0.038 & 0.038 & 0.025 & 0.026 & 0.0092 & 0.0095\\
\multirow{-2}{*}{\shortstack[c]{Primordial \\ non Gaussianities}} 
& $\Delta f_{NL}$ & 470 & 470 & 310 & 310 & 110 & 110\\ \hline
\end{tabular}

\label{tab:discuss_errors}
\tablefoot{Comparison of the impact of measurement errors on the main surveys discussed in this study. All analyses presented here use the full z--CR--HR information, and ten bins in redshift. eRASS:8 is All-Sky. Measurement error models are rescaled according to the depth of each survey.}
\end{table*}

\par We then compare the Fisher forecasts with the measurement errors included with the base case, for both surveys. Results are shown in Table \ref{tab:discuss_errors}. For all cases, the impact of measurement errors is almost negligible. This validates our initial approach, but also highlights a strength of \textit{ATHENA}: errors will be small enough to conserve all the informative content of its surveys. The same method can be used to model measurement errors on eRASS:8 X-ray observable diagrams and study there impact on the forecasted constraints. The results are presented in the rightmost columns of Table \ref{tab:discuss_errors}. We report no significant alteration of the constraints for $w$CDM and primordial non-Gaussianities; however, for $w_z$CDM, measurement errors increase $\Delta w_0$ ($\Delta w_a$) by 15\% (13\%). We anticipate that this effect will be accentuated for an eRASS:4 survey with only half the exposure time.

\subsection{Impact of priors}

\begin{table}[]
\caption{Survey B constraints for different prior sets.}
\centering
\begin{tabular}{c|c|ccc}
\hline
Case & 
Param. & Base & Planck$\times$4 & XXL/4 \\ \hline \hline
& $\Delta \Omega_m$ & 0.0094 & 0.010 & 0.0093 \\
& $\Delta \sigma_8$ & 0.0060 & 0.0062 & 0.0060\\
\multirow{-3}{*}{$w$CDM}   
& $\Delta w_0$      & 0.054 & 0.054 & 0.054\\ \hline

& $\Delta \Omega_m$ & 0.016 & 0.017 & 0.016 \\
& $\Delta \sigma_8$ & 0.0091 & 0.0092 & 0.0089 \\
& $\Delta w_0$      & 0.22 & 0.22 & 0.21 \\
\multirow{-4}{*}{$w_z$CDM}       
& $\Delta w_a$      & 0.87 & 0.87 & 0.85 \\ \hline

& $\Delta \Omega_m$ & 0.0094 & 0.0098 & 0.0094 \\
& $\Delta \sigma_8$ & 0.025 & 0.026 & 0.025 \\
\multirow{-2}{*}{\shortstack[c]{Primordial \\ non Gaussianities}} 
& $\Delta f_{NL}$ & 310 & 310 & 310 \\ \hline
\end{tabular}
\label{tab:discuss_priors}
\end{table}

Two arguments could be made about our assumptions on priors. Firstly, the use of Planck priors is highly constraining and may be questioned for instance if the Hubble tension remains in the 2030s. Secondly, the XXL scaling relation priors are too loose: given that the number of X-ray-detected clusters in the 2030s will largely exceed the current population, we  expect much better constraints on the scaling-relation parameters.
\par We investigated how changes to our base priors affect the forecasts, taking the case of survey B as an example. As a first test, we applied Planck priors broadened by a factor of 4. In a second independent test, we strengthened the XXL priors on the normalisation and slope of L--M and T--M by a factor of 4. The results are reported in Table \ref{tab:discuss_priors}. For $w_z$CDM, the constraints are only slightly broadened when $h$, $\Omega_b,$ and $n_s$ are given more latitude. The worst case is for $\Delta\sigma_8$, which increases by $\sim$20\%, but more importantly $\Delta w_0$ and $\Delta w_a$ are almost unchanged. On the contrary, with tighter priors on the scaling relations, we can expect a significant improvement on the $w_z$CDM results: 24\% (21\%) for $\Delta w_0$ ($\Delta w_a$). However, for primordial non-Gaussianities and $w$CDM, we observe no significant difference for these scenarios with respect to the base analysis case. This is a further indication that the broad mass and redshift range of the \textit{ATHENA} cluster samples allows a self-calibration of the scaling relations.

\subsection{Impact of spectroscopic redshifts}

In this section, we consider an optimistic scenario, where spectroscopic redshifts are available for both surveys. Number count cosmology can be improved with finer redshift bins in the analysis. The results presented in section \ref{sec:results} use large redshift bins of $\Delta z \sim 0.2$, which is consistent with the use of photometric redshifts. This working assumption is supported by the idea that the larger the sample and higher the redshift, the more difficult and time expensive it is to obtain redshifts through dedicated optical follow up. However, the availability of new spectrographs put in service by the launch of \textit{ATHENA} (e.g. \textit{4MOST}, \textit{ELT/MOSAIC}) could ease the redshift measurements for surveys A and B. Therefore, we consider in this section that we have access to spectroscopic redshifts for each object, and divide the redshift bin size by a factor of 4: $\Delta z \sim 0.05$. We compare this analysis to the survey B base case, for $w$CDM, $w_z$CDM, and primordial non-Gaussianities in Table \ref{tab:discuss_spectroz}. For $w$CDM and primordial non-Gaussianities, the spectroscopic redshifts do not show a major improvement on the constraints. However, in the $w_z$CDM case, there is a significant effect: $\Delta w_0$ is reduced by almost 30\% and $\Delta w_a$ by 20\%. 
\begin{table}[]
\centering
\caption{Survey B constraints with and without available spectroscopic redshifts.}
\begin{tabular}{c|c|ccc}
\hline
Case & 
Param. & Base & Spectro-$z$ \\ \hline \hline
& $\Delta \Omega_m$ & 0.0094 & 0.0087 \\
& $\Delta \sigma_8$ & 0.0060 & 0.0056 \\
\multirow{-3}{*}{$w$CDM}       
& $\Delta w_0$      & 0.54 & 0.05 \\ \hline

& $\Delta \Omega_m$ & 0.016 & 0.012 \\
& $\Delta \sigma_8$ & 0.0091 & 0.0071 \\
& $\Delta w_0$      & 0.22 & 0.16 \\
\multirow{-4}{*}{$w_z$CDM}       
& $\Delta w_a$      & 0.87 & 0.69 \\ \hline

& $\Delta \Omega_m$ & 0.0094 & 0.0092 \\
& $\Delta \sigma_8$ & 0.025 & 0.025 \\
\multirow{-2}{*}{\shortstack[c]{Primordial \\ non-Gaussianities}} 
& $\Delta f_{NL}$ & 310 & 300 \\ \hline
\end{tabular}
\label{tab:discuss_spectroz}
\end{table}

\subsection{Metallicity of the ICM}
 We assume a constant metallicity $Z=0.3Z_\odot$ in our modelling, which is a common choice in the literature. Metallicity has an effect on the shape of the X-ray observable diagrams, and a fixed value is a simplistic assumption. At the XMM spectral resolution, the effect of temperature is somewhat degenerate with that of metallicity, resulting in the so-called iron bias \citep{gastaldello_apparent_2010}. Hence, the impact of metallicity on the  XOD shape would be rather difficult to quantify, all the more so since the number of cluster photons can be as low as $\sim$~ 100. Moreover, no statistically significant observational constraints exist on the metallicity of high-z clusters. Simulations could provide insights with which to study this question, but the currently available results show discrepancies \citep[see e.g.][]{vogelsberger_uniformity_2018,pearce_redshift_2021}.

\section{Summary and conclusion} \label{sec:ccl}
We studied the potential of future deep X-ray surveys to constrain cosmology. We defined two surveys (A, 50 deg$^2$ at 80ks; and B, 200 deg$^2$ at 20ks) to be carried out by a modern and sensitive imager with a large FoV and a large collecting area, such as the \textit{ATHENA}/WFI project. We modelled the cluster selection function by requiring an S/N limit of 5 in a fixed optimised detection cell, and deduced the corresponding cluster number counts for both survey configurations. We then performed a cosmological Fisher analysis  based on the forward modelling of the distribution of the CR, HR, and $z$ cluster values, which constitutes our summary statistics.
We focused on  cosmological parameters that should still be relevant  in the late 2030s, namely the $w_z$CDM model, and local primordial non-Gaussianities. We summarise our main results below:
\begin{itemize}
    \item Surveys A and B are expected to detect some 5000 and  10000 clusters, respectively, in the [0.05 -- 2] redshift range. Both surveys will systematically detect hundreds of low-mass systems at $z>1$ down to a few $10^{13} h^{-1}$M$_\odot$, a population that is poorly characterised at present.
    \item Thanks to its larger number of clusters, Survey B has a greater constraining power, for both $w_z$CDM and local primordial non-Gaussianities.
    \item High-$z$ clusters play a major role in the obtained constraints; although they represent a small fraction ($\sim$15\%) of the total samples, they reduce the degeneracy between parameters, improving    $\Delta w_a$ by a factor of  ~2.3.  This is a remarkable result and paves the way for further prospective studies in correlation with the future S-Z (e.g. CMB-S4) and radio (e.g. SKA) observatories.
    \item The  $f_{NL}$ analysis shows the same trend: the constraint on $\Delta f_{NL}$ improves by a factor of $\sim$2.6. However, number counts alone do not provide competitive constraints on local primordial non-Gaussianities. In a future study, we shall address the impact of spatial clustering with the same  survey data, in which case the $f_{NL}$ constraints are expected to be several times stronger.
    \item The strength of our analysis lies in our forward modelling  of the z--CR--HR summary statistics, which bypasses the calculation of the individual cluster masses. Moreover, our approach allows the inclusion of all clusters down to the detection limit (not only e.g. those for which it is possible to determine the temperature). Dealing with deep surveys minimises errors on CR and HR for a large fraction of the cluster population. 
    \item Our results are robust with respect to the input priors on cluster physics. Indeed, the number of clusters to be detected both below and above $z=1$ is greater than the current samples used to derived scaling relations  by nearly two orders of magnitude. Our study shows that \textit{ATHENA} deep surveys have the capability to self-calibrate  the scaling relations while performing the cosmological analysis \citep{majumdar_self-calibration_2004}. Compared to \textit{eROSITA}, for which the detection limit was set to 50 photons, survey A (B) will yield at least $\sim$120 ($\sim$80) photons per cluster.
    \item The introduction of measurement errors has only a marginal effect on constraints yielded by \textit{ATHENA} surveys. However, such errors have to be accounted for in the case of \textit{eRASS:8} for the study of the DE EoS.
    \item Similarly, the availability of spectroscopic redshifts is not important when considering a $w$CDM cosmology or local primordial non-Gaussianities. However, it helps to further  disentangle $w_a$ and $w_0$ in a $w_z$CDM scenario.
    \item The present study highlights the impact of the $1<z<2$ range for a few currently debated cosmological parameters; we anticipate that it will also be relevant for new challenges that may arise between now and the \textit{ATHENA} mission.
\end{itemize}

This paper is a first attempt to estimate the cosmological potential of the high-redshift  (out to $z=2$) and low-mass cluster population for cosmology. Our results are promising and bring new scientific motivation for the \textit{ATHENA} mission. While the  5'' HEW PSF requirement is not essential for this work, it will nevertheless ease cluster detection and characterisation. \textit{LYNX} \citep{gaskin_lynx_2019}, a mission concept promoted by the 2020 Decadal Survey, would be another very promising X-ray observatory for accessing the high-$z$ clusters, with its High Definition X-ray Imager.\\
\par The survey characteristics (area, optimal aperture, and limiting count rate within the optimal aperture) needed to reproduce the selection functions are provided in Table \ref{tab:surveys}. Covariance matrices corresponding to each analysis case are available upon request to the authors. 

\begin{acknowledgements}
This work was supported by the Data Intelligence Institute of Paris (diiP), and IdEx Université de Paris (ANR-18-IDEX-0001). The authors thank Jean-Luc Sauvageot for useful considerations on the PSF, and François Lanusse and Valeria Pettorino for their valuable conversations about the Fisher analysis.
\end{acknowledgements}

\bibliographystyle{aa}
\bibliography{athena_paper_cosmo}

\begin{appendix}
\section{Derivation of the non-Gaussian HMF}\label{app:HMFnongauss}

We provide step-by-step details of our derivation of Eq. \ref{eq:deltaR3} from Eq. \ref{eq:bardeen}. We take the Fourier transform in the form of:
\begin{equation}
    \delta(\Vec{x}) = \int \frac{d\Vec{k}}{(2 \pi)^3} e^{-i\Vec{k}.\Vec{x}} \delta(\Vec{k})
    \label{eq:fourier}
.\end{equation} 
The power spectrum of the matter density fluctuations is defined as:
\begin{equation}
    \langle \delta(\Vec{k_1}) \delta(\Vec{k_2}) \rangle
    = \delta_D(\Vec{k_1}+\Vec{k_2}) P(k_1) (2 \pi)^3
    \label{eq:Pk}
.\end{equation}
From equation \ref{eq:fourier}, we can express $\delta_R$, the smoothed density fluctuations field on the scale $R$, in the Fourier space as:
\begin{equation}
    \delta_R(\Vec{k}) = \delta(\Vec{k}) W_R(k)
    \label{eq:deltaWR}
.\end{equation}

Also, we note that equation \ref{eq:alphaphi} gives
\begin{equation}
    P(k) = \alpha(k,z)^2 P_{\Phi \Phi}
.\end{equation}
Using equation \ref{eq:bardeen}, removing second-order terms in $f_{NL}$, we also get
\begin{equation}
    P(k) = \alpha(k,z)^2 P_{\phi \phi}
    \label{eq:alphaPphi}
.\end{equation}

Before expressing $\langle \delta_R^3 \rangle$, we will look for the expression of $\langle \Phi(\Vec{k_1}) \Phi(\Vec{k_2}) \Phi(\Vec{k_3}) \rangle$. We only conserve the first-order terms in $f_{NL}$, and as they are symmetric, we can write
\begin{equation}
\begin{split}
    &\langle \Phi(\Vec{k_1}) \Phi(\Vec{k_2}) \Phi(\Vec{k_3}) \rangle \\
    &\ = 3 f_{NL} \langle
    \int \frac{d\Vec{k_1'}d\Vec{k_1''}}{(2 \pi)^3}
    \delta_D(\Vec{k_1'}+\Vec{k_1^{\prime\prime}} -\Vec{k_1})
    \phi(\Vec{k_1'})\phi(\Vec{k_1''})\phi(\Vec{k_2})\phi(\Vec{k_3})
    \rangle
\end{split}
.\end{equation}

Then, using Wick's theorem, we have
\begin{equation}
\begin{split}
     \langle \Phi(\Vec{k_1}) \Phi(\Vec{k_2}) \Phi(\Vec{k_3}) \rangle 
     \ = & 3 f_{NL} \int \frac{d\Vec{k_1'}d\Vec{k_1''}}{(2 \pi)^3}
    \delta_D(\Vec{k_1'}+\Vec{k_1''} -\Vec{k_1}) \\
    & \times( \langle \phi(\Vec{k_1'})\phi(\Vec{k_1''}) \rangle
    \langle \phi(\Vec{k_2})\phi(\Vec{k_3}) \rangle \\
    & \ + \langle \phi(\Vec{k_1'})\phi(\Vec{k_2}) \rangle
    \langle \phi(\Vec{k_1''})\phi(\Vec{k_3}) \rangle \\
    & \ + \langle \phi(\Vec{k_1'})\phi(\Vec{k_3}) \rangle
    \langle \phi(\Vec{k_1''})\phi(\Vec{k_2}) \rangle )
\end{split}
.\end{equation}

We then express $\langle \delta_R^3 \rangle $ as
\begin{equation}
    \begin{split}
        \langle \delta_R^3 \rangle = 
        \int &\frac{d\Vec{k_1}d\Vec{k_2}d\Vec{k_3}}{(2 \pi)^{9}} e^{-i(\Vec{k_1}+\Vec{k_2}+\Vec{k_3}).\Vec{x}} \\
        & \times W_R(k_1) W_R(k_2) W_R(k_3)  \alpha(k_1,z) \alpha(k_2,z) \alpha(k_3,z) \\
        &\times \langle \Phi(\Vec{k_1}) \Phi(\Vec{k_2}) \Phi(\Vec{k_3}) \rangle
    \end{split}
.\end{equation}

Noticing that, for $\Vec{k_1}=0$, we have $\alpha(k_1)=0$, the first term in Wick's theorem brings no contribution. The remaining two terms are symmetric, and so we can write

\begin{equation}
    \begin{split}
        \langle \delta_R^3 \rangle = 
        \int &\frac{d\Vec{k_1}d\Vec{k_2}d\Vec{k_3}}{(2 \pi)^{9}} e^{-i(\Vec{k_1}+\Vec{k_2}+\Vec{k_3}).\Vec{x}} \\
        &\times W_R(k_1) W_R(k_2) W_R(k_3)  \alpha(k_1,z) \alpha(k_2,z) \alpha(k_3,z) \\
        &\times 3 f_{NL} \int \frac{d\Vec{k_1'}d\Vec{k_1''}}{(2 \pi)^3}
        \delta_D(\Vec{k_1'}+\Vec{k_1^{\prime\prime}} -\Vec{k_1}) \\
        &\times \left[ 2 \langle \phi(\Vec{k_1'})\phi(\Vec{k_2}) \rangle
        \langle \phi(\Vec{k_1''})\phi(\Vec{k_3}) \rangle \right]
\end{split}
.\end{equation}

Then, using equation \ref{eq:Pk}, we can replace the terms in the last line and reorganise the expression to get:
\begin{equation}
    \begin{split}
        \langle \delta_R^3 \rangle = 
         6 f_{NL} \int &\frac{d\Vec{k_1}d\Vec{k_2}d\Vec{k_3}d\Vec{k_1'}d\Vec{k_1''}}{(2 \pi)^{12}} e^{-i(\Vec{k_1}+\Vec{k_2}+\Vec{k_3}).\Vec{x}} \\
        &\times W_R(k_1) W_R(k_2) W_R(k_3)  \alpha(k_1,z) \alpha(k_2,z) \alpha(k_3,z) \\
        & \times \delta_D(\Vec{k_1'}+\Vec{k_1^{\prime\prime}} -\Vec{k_1})
        \delta_D(\Vec{k_1'}+\Vec{k_2}) P_{\phi \phi}(k_2)  \\
        & \times \delta_D(\Vec{k_1''}+\Vec{k_3}) P_{\phi \phi}(k_3) (2 \pi)^6
    \end{split}
.\end{equation}

The Dirac functions give $\Vec{k_2} = -\Vec{k_1'}$, $\Vec{k_3} = -\Vec{k_1''}$ and also $\Vec{k_1} + \Vec{k_2} = -\Vec{k_3}$. The expression is then simplified:
\begin{equation}
    \begin{split}
        \langle \delta_R^3 \rangle = 
         6 f_{NL} \int &\frac{d\Vec{k_2}d\Vec{k_3}}{(2 \pi)^{6}}  W_R(k_1) W_R(k_2) W_R(k_3) \\
        &\times \alpha(k_1,z) \alpha(k_2,z) \alpha(k_3,z) P_{\phi \phi}(k_2) P_{\phi \phi}(k_3)
    \end{split}
.\end{equation}

We then rename  the vectors: $\Vec{k_2}$ is renamed $\Vec{k_1}$,  $\Vec{k_3}$ becomes $\Vec{k_2}$, and $\Vec{k_1}$ becomes  $\Vec{k_{12}} = -\Vec{k_3} = \Vec{k_1} + \Vec{k_2}$. We use equation \ref{eq:alphaPphi} to write

\begin{equation}
    \begin{split}
        \langle \delta_R^3 \rangle = 
         6 f_{NL} \int &\frac{d\Vec{k_1}d\Vec{k_2}}{(2 \pi)^{6}}  W_R(k_1) W_R(k_2) W_R(k_{12}) \\
        &\times \alpha(k_1,z) \alpha(k_2,z) \alpha(k_{12},z)
        \frac{P(k_1) P(k_2)}{\alpha(k_1,z)^2 \alpha(k_2,z)^2}
    \end{split}
,\end{equation}
and so
\begin{equation}
    \begin{split}
        \langle \delta_R^3 \rangle = 
         6 f_{NL} \int &\frac{d\Vec{k_1}d\Vec{k_2}}{(2 \pi)^{6}}  W_R(k_1) W_R(k_2) W_R(k_{12}) \\
        & \times \frac{\alpha(k_{12},z)}{\alpha(k_1,z) \alpha(k_2,z)}P(k_1) P(k_2)
    \end{split}
.\end{equation}
Finally, we use the fact that $k_{12}^2=k_1^2 + k_2^2 + 2\mu k_1k_2$ to write the final expression:
\begin{equation}
    \begin{split}
        \langle \delta_R^3 \rangle \ = \ & 6 f_{NL} \frac{8\pi^2}{(2\pi)^6}
         \int_0^{\infty}dk_1 k_1^2 \int_0^{\infty}dk_2 k_2^2 \\
         & \times \int_{-1}^{1}d\mu W_R(k_1)W_R(k_2)W_R(k_{12})
         \frac{\alpha(k_{12})}{\alpha(k_{1})\alpha(k_{2})}P(k_1)P(k_2)
  \end{split} 
.\end{equation}

\section{Impact of high-$z$ clusters in survey A}\label{app:highzA}

\begin{figure*}
   \centering
   \includegraphics[width=15cm]{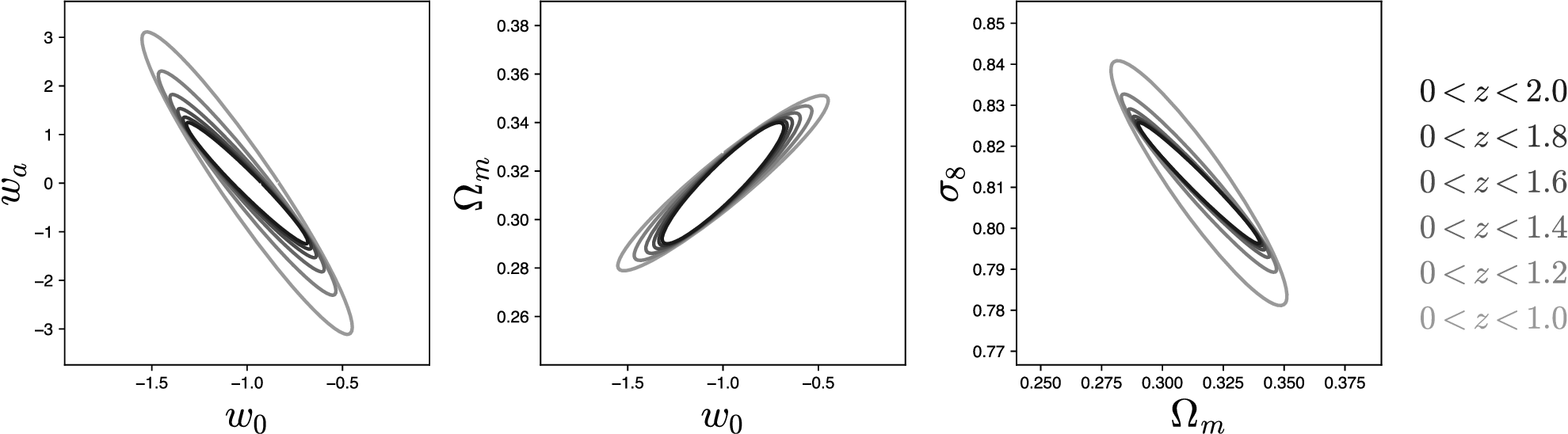}
      \caption{Impact of $z>1$ clusters in survey A on $w_0, w_a$}
         \label{fig:w0wa_slices_A}
\end{figure*}

\begin{figure}
   \centering
   \includegraphics[width=7cm]{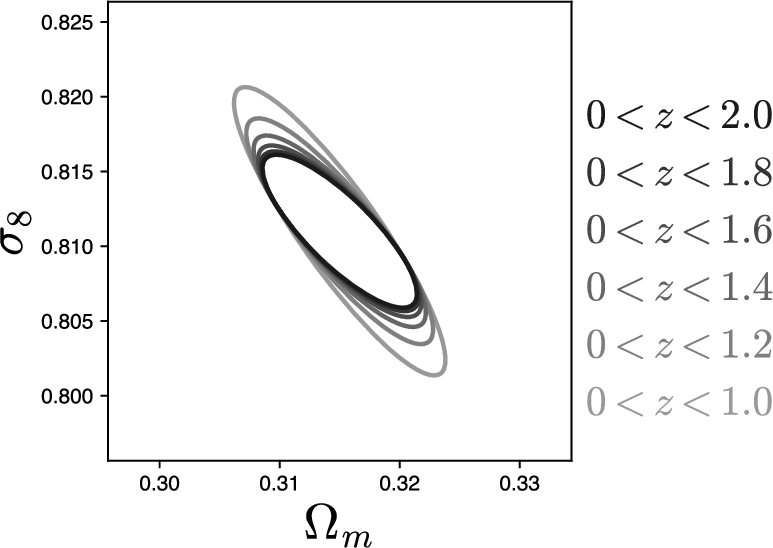}
      \caption{Impact of $z>1$ clusters in survey A on $\Omega_m, \sigma_8$}
         \label{fig:lcdm_slices_A}
\end{figure}

\begin{figure}
   \centering
   \includegraphics[width=8cm]{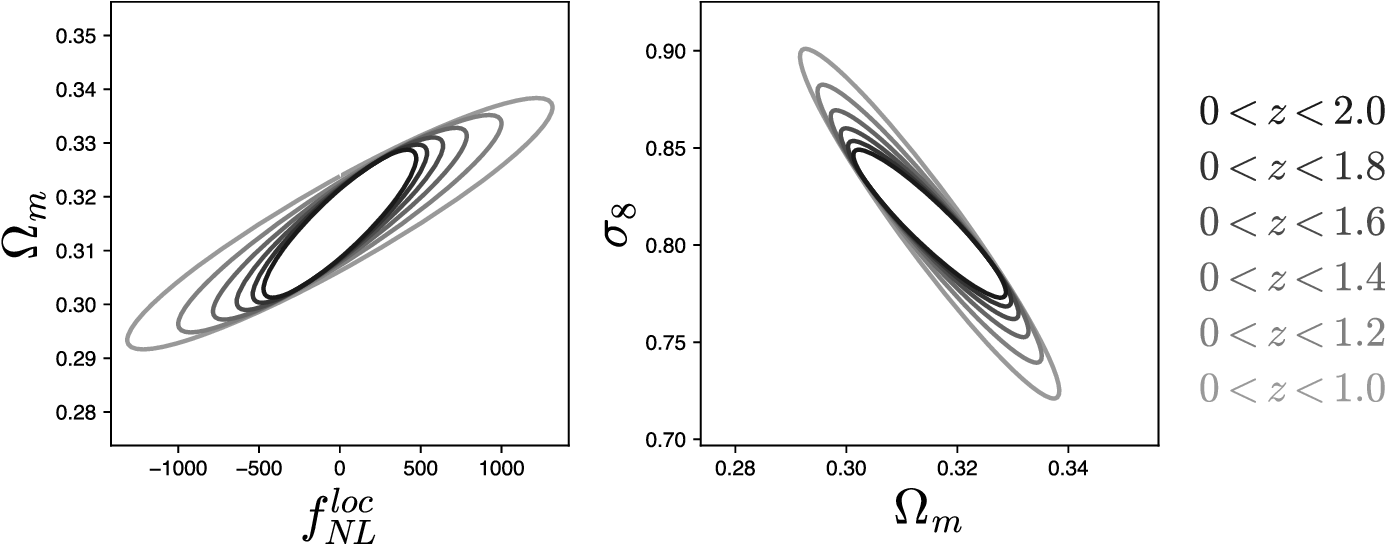}
      \caption{Impact of $z>1$ clusters in survey A on $f_{NL}$}
         \label{fig:fNL_slices_A}
\end{figure}

We discuss here the impact of survey A high-redshift clusters on the cosmological analysis. Similarly to figures \ref{fig:lcdm_slices_B}, \ref{fig:w0wa_slices_B}, \ref{fig:fNL_slices_B}, we perform successive truncated analysis to understand the contribution of the high-redshift systems. The figure \ref{fig:lcdm_slices_A} shows the evolution of forecasted constraints in the vanilla $\Lambda$CDM case, figure \ref{fig:w0wa_slices_A} in the $w_z$CDM cosmology, and figure \ref{fig:fNL_slices_A} for local primordial non-Gaussianities. In all these figures, we observe the very same statistical phenomenon as in Fig. \ref{sec:results}: high-redshift clusters tilt the ellipses and hence break the degeneracy between parameters. The more degenerate the parameters, the stronger the effect: in \ref{fig:lcdm_slices_A}, $\Delta \sigma_8$ is reduced by a factor of 1.9; in \ref{fig:w0wa_slices_A}, $\Delta w_a$ is reduced by a factor of 2.5, and in \ref{fig:fNL_slices_A},  $\Delta f_{NL}$ shrinks by a factor of 2.8. We note that the effect is slightly stronger for survey A than for survey B: this could be due to the fact that (i) in survey A the cluster populations for $z<1$ and $z>1$ are more balanced than in survey B, and (ii) survey A probes the HMF to lower masses.

\end{appendix}
\end{document}